\documentclass[12pt]{iopart}
\usepackage{graphicx}
\def\Re{{\cal R \mskip-4mu \lower.1ex \hbox{\it e}\,}}
\def\Im{{\cal I \mskip-5mu \lower.1ex \hbox{\it m}\,}}
\def\ie{{\it i.e.}}
\def\eg{{\it e.g.}}

\def\sub#1{_{\lower.25ex\hbox{$\scriptstyle#1$}}}
\def\tev{\,{\ifmmode\mathrm {TeV}\else TeV\fi}}
\def\gev{\,{\ifmmode\mathrm {GeV}\else GeV\fi}}
\def\mev{\,{\ifmmode\mathrm {MeV}\else MeV\fi}}
\def\mpl{\ifmmode M_{pl}\else $M_{pl}$\fi}
\def\mpl{\ifmmode \overline M_{Pl}\else $\bar M_{Pl}$\fi}
\def\to{\rightarrow}

\def\subw{_{\rm w}}
\def\mh{\ifmmode m\sbl H \else $m\sbl H$\fi}
\def\mch{\ifmmode m_{H^\pm} \else $m_{H^\pm}$\fi}
\def\mt{\ifmmode m_t\else $m_t$\fi}
\def\mc{\ifmmode m_c\else $m_c$\fi}
\def\mz{\ifmmode M_Z\else $M_Z$\fi}
\def\mw{\ifmmode M_W\else $M_W$\fi}
\def\mws{\ifmmode M_W^2 \else $M_W^2$\fi}
\def\mhs{\ifmmode m_H^2 \else $m_H^2$\fi}   
\def\mzs{\ifmmode M_Z^2 \else $M_Z^2$\fi}
\def\mts{\ifmmode m_t^2 \else $m_t^2$\fi}
\def\mcs{\ifmmode m_c^2 \else $m_c^2$\fi}
\def\mchs{\ifmmode m_{H^\pm}^2 \else $m_{H^\pm}^2$\fi}
\def\ztwo{\ifmmode Z_2\else $Z_2$\fi}
\def\zone{\ifmmode Z_1\else $Z_1$\fi}
\def\mtwo{\ifmmode M_2\else $M_2$\fi}
\def\mone{\ifmmode M_1\else $M_1$\fi}
\def\tb{\ifmmode \tan\beta \else $\tan\beta$\fi}
\def\xw{\ifmmode x\subw\else $x\subw$\fi}
\def\ch{\ifmmode H^\pm \else $H^\pm$\fi}
\def\lum{\ifmmode {\cal L}\else ${\cal L}$\fi}
\def\inpb{\,{\ifmmode {\mathrm {pb}}^{-1}\else ${\mathrm {pb}}^{-1}$\fi}}
\def\infb{\,{\ifmmode {\mathrm {fb}}^{-1}\else ${\mathrm {fb}}^{-1}$\fi}}
\def\epem{\ifmmode e^+e^-\else $e^+e^-$\fi}
\def\ppb{\ifmmode \bar pp\else $\bar pp$\fi}
\def\bsg{\ifmmode B\to X_s\gamma\else $B\to X_s\gamma$\fi}
\def\bsll{\ifmmode B\to X_s\ell^+\ell^-\else $B\to X_s\ell^+\ell^-$\fi}
\def\bstt{\ifmmode B\to X_s\tau^+\tau^-\else $B\to X_s\tau^+\tau^-$\fi}
\def\lamt{\ifmmode \tilde\lambda\else $\tilde\lambda$\fi}
\def\shat{\ifmmode \hat s\else $\hat s$\fi}
\def\that{\ifmmode \hat t\else $\hat t$\fi}
\def\uhat{\ifmmode \hat u\else $\hat u$\fi}

\newskip\zatskip \zatskip=0pt plus0pt minus0pt
\def\matth{\mathsurround=0pt}
\def\lsim{\mathrel{\mathpalette\atversim<}}
\def\gsim{\mathrel{\mathpalette\atversim>}}
\def\atversim#1#2{\lower0.7ex\vbox{\baselineskip\zatskip\lineskip\zatskip
  \lineskiplimit 0pt\ialign{$\matth#1\hfil##\hfil$\crcr#2\crcr\sim\crcr}}}

\def\grtsim{\,\,\rlap{\raise 3pt\hbox{$>$}}{\lower 3pt\hbox{$\sim$}}\,\,}
\def\lsim{\,\,\rlap{\raise 3pt\hbox{$<$}}{\lower 3pt\hbox{$\sim$}}\,\,}

\begin{document}
\begin{titlepage}
\rightline{\vbox{\halign{&#\hfil\cr
&SLAC-PUB-13562\cr
}}}

\title{Dark Matter in the MSSM}

\author{R C Cotta, J S Gainer, J L Hewett and T G Rizzo}
\address{
  SLAC National Accelerator Laboratory, 2575 Sand Hill Rd., Menlo
  Park, CA, 94025, USA
}
\ead{rcotta@stanford.edu, jgainer@slac.stanford.edu, 
hewett@slac.stanford.edu, rizzo@slac.stanford.edu}
\begin{abstract}
We have recently examined a large number of points in the parameter
space of the phenomenological MSSM, the 19-dimensional parameter space
of the CP-conserving MSSM with Minimal Flavor Violation.  We
determined whether each of these points satisfied existing
experimental and theoretical constraints.  This analysis provides
insight into general features of the MSSM without reference to a
particular SUSY breaking scenario or any other assumptions at the GUT
scale. This study opens up new possibilities for SUSY
phenomenology at colliders as well as in both direct and 
indirect detection searches for dark matter. 
\end{abstract}

\submitto{\NJP}
\end{titlepage}

\section{Introduction} 

Supersymmetry (SUSY) represents an appealing possibility for Beyond
the Standard Model Physics; its discovery would help provide answers
to many of the preeminent questions in particle physics, astrophysics,
and cosmology.
However, as no sparticles have been observed, it is clear that if
SUSY exists, it must be broken. 
The mechanism which could break SUSY is a question of great
importance, and there is an ever growing list of possible scenarios,
including mSUGRA\cite{msugrab}, GMSB\cite{gmsbb}, AMSB\cite{amsbb},
and gaugino mediated supersymmetry breaking\cite{ggmsbb}.  
In each of these scenarios, the SUSY spectrum is described by a
handful of parameters, generally defined at the SUSY breaking scale;
straightforward RGE running of sparticle masses and coupling constants
yields predictions for the mass spectra and decay patterns of the
various sparticles at energy scales relevant for colliders or
cosmology.  However, these SUSY breaking scenarios are restrictive and
predict specific phenomenologies for colliders and cosmology that
do not represent the full range of possible SUSY signatures.

It is clearly desirable to study the MSSM more broadly without making
simplifying assumptions at the high scale that may turn out to be
unwarranted.  However, the MSSM requires 105 parameters to describe
SUSY breaking, in addition to the parameters of the SM\cite{mssmrev}.
Obviously this is far too many parameters to study directly, so some
simplifying assumptions must be made.  Here we will restrict ourselves
to the CP-conserving MSSM (\ie, no new phases) with minimal flavor
violation (MFV)\cite{mfv}.  Additionally, we require that the first two
generations of sfermions be degenerate as motivated by constraints
from flavor physics. We are then left with 19  independent, real,
weak-scale, SUSY Lagrangian parameters, namely the gaugino masses
$M_{1,2,3}$, the Higgsino mixing parameter $\mu$, the ratio of the
Higgs vevs $\tan \beta$, the mass of the pseudoscalar Higgs boson
$m_A$, and the 10 squared masses of the sfermions ($m_{\tilde{q}1,3}$,
$m_{\tilde{u}1,3}$,$m_{\tilde{d}1,3}$,$m_{\tilde{l}1,3}$,and
$m_{\tilde{e}1,3}$).  We include independent $A$-terms only for
the third generation ($A_b$, $A_t$, and $A_\tau$) due to the small
Yukawa couplings for the first two generations.  This set of 19
parameters has been called the phenomenological MSSM
(pMSSM){\cite{Djouadi:2002ze}}.

To study the pMSSM, we performed a scan over this 19-dimensional
parameter space assuming flat priors for the specified ranges\cite{Berger:2008cq}:
\begin{eqnarray}
100 \gev \leq m_{\tilde f} \leq 1\tev \,, \nonumber\\
50\gev \leq |M_{1,2},\mu|\leq 1 \tev\,, \nonumber \\ 
100 \gev \leq M_3\leq 1 \tev\,, \nonumber \\ 
|A_{b,t,\tau}| \leq 1 \tev\,, \\
1 \leq \tan \beta \leq 50\,, \nonumber \\ 
43.5\gev \leq m_A \leq 1 \tev\,. \nonumber  
\end{eqnarray}

The value of 43.5 GeV in the last constraint was chosen to avoid the possible on-shell 
decay $Z\to hA$. 
We randomly generated $10^7$ points in this parameter space and subjected them to
a number of existing theoretical and experimental constraints.  We
also performed a scan with log priors and slightly different mass
ranges (that we will not employ here) in order 
to gauge the influence of priors on our results; we found that
these results are substantially similar to those obtained in our flat prior
scan\cite{Berger:2008cq}.
Using these parameters we generate a SUSY spectrum utilizing
SuSpect2.34\cite {Djouadi:2002ze}.  By convention, the parameters are
specified at the scale given by the geometric mean of the two stop
masses.  The input values for the SM parameters used in our 
analysis are given in~\cite{Berger:2008cq}.

We then apply a series of constraints obtaining a set of models that 
satisfy all existing theoretical and experimental data. (This is the 
so-called ``flat prior'' set obtained in Ref.{\cite {Berger:2008cq}}.) 
In the analysis below we will discuss the applied constraints then 
will examine the properties of the LSP at the parameter points which
remain viable.  In particular, we will examine the gaugino and
Higgsino content of the LSP (which is always the lightest neutralino).
We will also discuss the nature of the nLSP and the difference between
its mass and the LSP mass; this is important, for example, in
coannihilation processes.  We will then examine the signatures in
direct and indirect WIMP detection experiments obtained for these
points in parameter space.

\section{Theoretical and Experimental Constraints}

We now discuss the theoretical and experimental constraints which we
applied to the generated parameter space points (which we shall
hereafter refer to as ``models'' for convenience).  We will present
each of these briefly in turn; for more details, one should
consult~\cite{Berger:2008cq}.

\subsection{Theoretical Constraints}

We demand that the sparticle spectrum not have tachyons or color or
charge breaking (CCB) minima in the scalar potential.  
We also require that the Higgs potential be bounded from below and
that electroweak symmetry breaking be consistent. 
We assume that the LSP, which will be absolutely stable, be a
conventional thermal relic so that the LSP can be identified as 
the lightest neutralino.
If it is a significant component of the dark matter, the LSP must be
uncolored and uncharged, thus the LSP can only be a sneutrino or a
neutralino.
The possibility that the LSP is a sneutrino can be easily eliminated
in the pMSSM by combining several of the experimental constraints,
particularly those involving direct detection of sneutrino WIMPs
and the invisible width of the $Z$, as discussed below.

\subsection{Low Energy Constraints}

The code  micrOMEGAs2.20{\cite {MICROMEGAS}} was used to evaluate the
following observables for each point in the parameter space:
$\Delta \rho$, the decay rates for $b\to s\gamma$ and $B_s \to
\mu^+\mu^-$, and the $g-2$ of the muon.  In addition, we evaluate the
branching fraction for $B\to \tau \nu$ following{\cite{gino}}
and {\cite {ems}}.  The ranges that we allow for these observables are
listed in Table~\ref{low energy table}.  
The large range for the SUSY
contribution to $g-2$ ($\sim 6 \sigma$) is due to the evolving discrepancy
between theory and experiment{\cite{Bennett:2006fi}}.

\begin{table}
\centering
\begin{tabular}{|c|c|c|} \hline\hline 
Constraint & Range & References \\ \hline
$\Delta \rho$          & $-0.0007$ - $0.0026$ & {\cite{Amsler:2008zz}} \\
$b \to s \gamma$       & $2.5\times10^{-4}$ - $4.1\times10^{-4}$
&{\cite {HFAG}}{\cite{Misiak:2006zs}}{\cite{Becher:2006pu}} \\
$B_s \to \mu^+ \mu^-$  & 0 - $4.5\times10^{-8}$ &{\cite{toback}} \\
$\Delta_{\mathrm{SUSY}}(g-2)_\mu$ & $-1.0\times10^{-9}$ - $4.0 \times 10^{-9}$ &  
{\cite{Bennett:2006fi}}{\cite{Passera:2008hj}}{\cite{DeRafael:2008iu}} \\
$B \to \tau \nu$       & $5.5\times 10^{-5}$ - $2.27\times 10^{-4}$&
{\cite {gino}}{\cite {ems}}{\cite {HFAG}}{\cite {Chang}}\\
\hline\hline
\end{tabular}
\caption{Ranges allowed for various low energy observables in our analysis.}
\label{low energy table}
\end{table}

We implemented constraints from meson-antimeson mixing{\cite{mesonmix}}
by assuming MFV{\cite{mfv}}, imposing first and second generation mass
degeneracy, and  demanding that the ratio of first/second and
third generation squark soft breaking masses (of a given flavor and
helicity) differ from unity by no more than a factor of $5$.  We
also imposed analogous restrictions in the slepton sector.

\subsection{LEP Constraints}

We now consider the constraints that arise from LEP data.  Due to
running LEP at the $Z$ pole, it is very unlikely that there can be charged
sparticles with masses below $M_Z/2$. The same constraint is applied
to the lightest neutral Higgs boson.  Data from LEPII{\cite
  {lepstable}} suggests that there are no new {\it stable} charged
particles of any kind with masses below 100 GeV.   We also require
that any new contributions to the
invisible width of the $Z$ boson be $\leq 2$ MeV{\cite {LEPEWWG}};
this constraint eliminates the possibility of certain species of
neutralinos having masses below $M_Z/2$.

Following ALEPH{\cite {ALEPH}} we implement a lower limit of $92$ GeV
on first and second generation squark masses, provided that the gluino
is more massive than the squarks and the mass difference ($\Delta m$)
between the squark and the LSP is $\geq 10$ GeV.  We also implement a
similar cut (following~\cite{bbb}) on the mass of sbottom quarks
requiring that their mass be greater than $95$ GeV (in addition to
$\Delta m \geq 10$ GeV, and the mass being less than the gluino mass).
The situation for stops is slightly more complex\cite{LEPSUSY}; we
demand that the lightest stop mass be greater than $97$ GeV if the
stop is too light to decay into $Wb\chi_1^0$.  If the stop can decay
to $\ell b\tilde \nu$; we have a lower limit of $95$ GeV on its mass.

Following~{\cite{LEPSUSY}}, we demand that right-handed sleptons have
masses greater than $100$, $95$, or $90$ GeV for selectrons, smuons,
and staus respectively.  We only apply this limit when the 
condition $0.97 m_{\mathrm{slepton}} > m_{\mathrm{LSP}}$ is satisfied.  
We can also apply these bounds to
left-handed sleptons, provided that the neutralino $t-$channel diagram
may be neglected in the case of selectrons; we assume that this is the
case.

We demand that chargino masses be greater than $103$ GeV, provided
that the LSP-chargino mass splitting is $\Delta
m > 2$ GeV\cite{LEPSUSY}.  
If $\Delta m<2$ GeV, the bound is $95$ GeV.  It should be
noted that when $\Delta m$ is very small ($\lsim 100 \mev$), the chargino is
stable for detector length scales and the model will be excluded by
the stable charged particle constraints.  In the case where the
lightest chargino is dominantly Wino, we can only apply this limit
when the electron sneutrino $t-$channel diagram is negligible; we take
this to be the case when the electron sneutrino is more massive than 160 GeV. 

The LEP Higgs Working Group{\cite {LEPHIGGS}},
provides five sets of constraints on the MSSM Higgs sector imposed by
LEPII data.  These are essentially limits on the Higgs-Z coupling times
the branching fraction for decay to given final states, as a function
of the Higgs masses. We employ SUSY-HIT\cite{SUSYHIT} to analyze these.
We include a theoretical uncertainty on the calculated mass of the
lightest Higgs
boson of approximately $3$ GeV{\cite {uncertain}} when applying these
constraints.

\subsection{Tevatron Constraints}

We also employ constraints from the Tevatron. 
We obtained restrictions on the squark and gluino sectors arising from
the null result of the D0 multijet plus missing energy search{\cite{domet}}. 
We generalize their analysis to render it model independent, by
generating multijet plus missing energy events for our model spectrum using
PYTHIA6.4{\cite {PYTHIA}} (which we provide with a
SUSY-HIT\cite{SUSYHIT} decay table) as interfaced to PGS4
{\cite  {PGS}}. 
PGS4 provides a fast detector simulation and is used to
impose the kinematic cuts used in the D0 analysis.  
We weigh our results with K factors computed using PROSPINO2.0
{\cite  {PROSPINO}}.
The $95\%$ CL upper limit on the number of signal events, as defined
by the D0 analysis, is $8.34$ (for the $2.1\infb$ data set considered)
using the method of Feldman and Cousins{\cite {fc}}.  
Analogously, we employ constraints from the CDF search for trileptons
plus missing energy{\cite {cdftrilepton}}, which we also generalize to
the full pMSSM. We only make use of the CDF `3 tight lepton' analysis
as it is the cleanest and easiest to implement with PGS4; we also use
a K-factor of $1.3$ for all models.  Here the $95\%$ CL
upper bound on the possible SUSY signal in the channel we are
considering is $4.65$ events for the luminosity of $2.02\infb$ used in
the CDF analysis.  

In addition to these collider signature bounds, we also employ
the experimental constraint{\cite {tevhiggs}} resulting from direct searches 
for the new Higgs fields in the MSSM: for the narrow mass range $90 \leq m_A \leq 100$,
$\tan \beta$ is restricted to the region $\tan \beta \geq 1.2m_A-70$. This range is 
excluded as the Tevatron would have otherwise discovered at least one of the heavier 
Higgs bosons.   
Also, D0{\cite {dostable}} has obtained lower limits on the mass of
heavy stable charged particles.  
We take this constraint to be $m_{\chi^+}\geq 206 |U_{1w}|^2 +171
|U_{1h}|^2$ GeV at $95\%$ CL for charginos, 
where the matrix entries $U_{1w}$ and
$U_{1h}$ determine the Wino/Higgsino content of the lightest chargino. 
We use this to interpolate between the separate Wino and
Higgsino results provided by D0. 

CDF and D0 also have analyses that search for light stops and
sbottoms{\cite {stops}} which include a number of assumptions about
the SUSY mass spectrum, sparticle decay channels, etc.  In general
they are only applicable when the sbottoms or stops are lighter than
the top quark.  These searches are difficult to implement
in a model-independent pMSSM context.  Thus we exclude models with
light ($m < m_t$) stops or sbottoms from our final set of models; this
only affected $\sim 1000$ models.
  
\subsection{Astrophysical Constraints}

There are two constraints from considering the LSP as a long-lived
relic.  As noted above, we demand that the LSP be the lightest
neutralino.  We also require, following the 5 year WMAP
measurement{\cite {Komatsu:2008hk}} of the relic density, that $\Omega
h^2|_{\mathrm{LSP}} \leq 0.121$.
In not employing a lower bound on $\Omega h^2|_{\mathrm{LSP}}$ for our
models, we acknowledge the possibility
that even within the MSSM and the thermal relic framework, dark matter
may have multiple components with the LSP being just one possible
contributor; we thus only require that the LSP not have a relic
density too large to be consistent with WMAP.
However, in discussing results below, we will also discuss a subset of
models for which $0.1 \leq \Omega h^2|_{LSP} \leq 0.121$; these
represent the more standard assumption that the LSP is the dominant,
perhaps only, component of the relic density.

We also obtain constraints from attempts to detect dark matter
directly{\cite{dmsearch}}.
Generally, the strongest constraints come from the spin-independent
WIMP-nucleon cross sections, hence we only implement bounds on our
models from these; inspection of the spin-dependent WIMP-nucleon cross
sections in our models confirms that this approach is reasonable.
Both spin-independent and spin-dependent cross sections were
calculated using micrOMEGAs2.21{\cite {MICROMEGAS}}.
We implement cross section limits from XENON10{\cite
  {XENON10}}, CDMS{\cite {CDMS}}, CRESST I{\cite {CRESST}} 
and DAMA{\cite {DAMA}} data. 
Since these cross sections depend on some low energy quantities for
which the uncertainties are relatively large (\eg,  nuclear form
factors), we do not exclude models with WIMP-nucleon spin-independent cross
sections as much as 4 times larger than the experimental bounds. It
should be noted that many of our models predict a value
$\Omega h^2|_{\mathrm{LSP}}$ which is less than that observed by WMAP
and supernova searches.  We thus scale our cross sections 
to take this into account.

\section{Results}

As noted above we randomly generated $10^7$ parameter space points (\ie,
models) in a 19-dimensional pMSSM parameter space using flat priors.
Only $\sim 68.5 \cdot 10^3$ of these models satisfy all the
constraints listed in the previous section.  The properties of these
models are described in much greater detail in~\cite{Berger:2008cq}.  Here
we will discuss the attributes of these models
which are most important astrophysically.  In particular we will
examine the mass and composition of the LSP and nLSPs, the predicted
relic density, as well as direct and indirect dark matter detection
signals from these models.

\subsection{LSP and nLSP}

\begin{figure}[htbp]
\begin{center}
\includegraphics[width=14.0cm,angle=0]{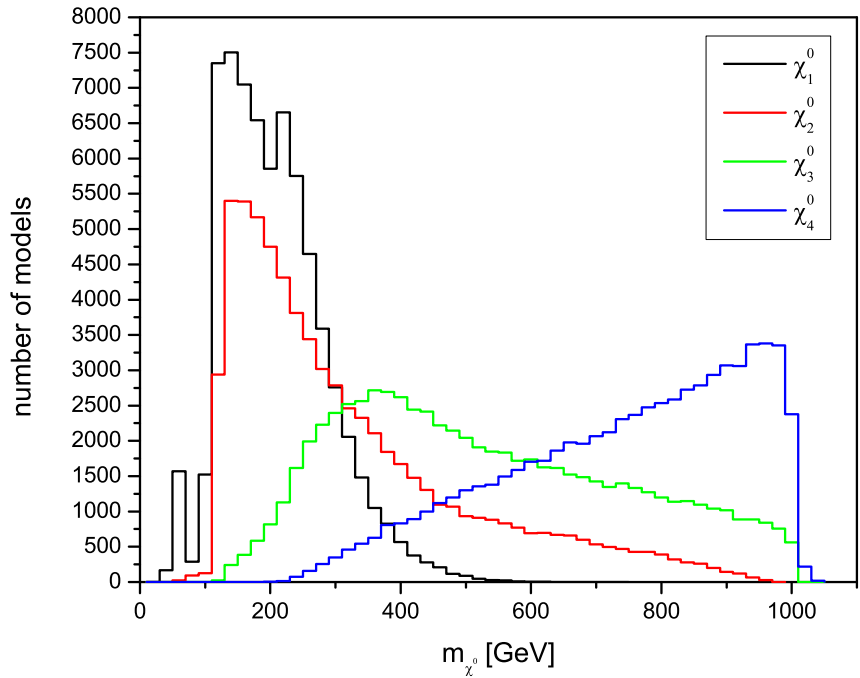}
\end{center}
\caption{Distribution of neutralino masses for our set of
  models.}
\label{fig1}
\end{figure}

\begin{figure}[htbp]
\begin{center}
\includegraphics[width=14.0cm,angle=0]{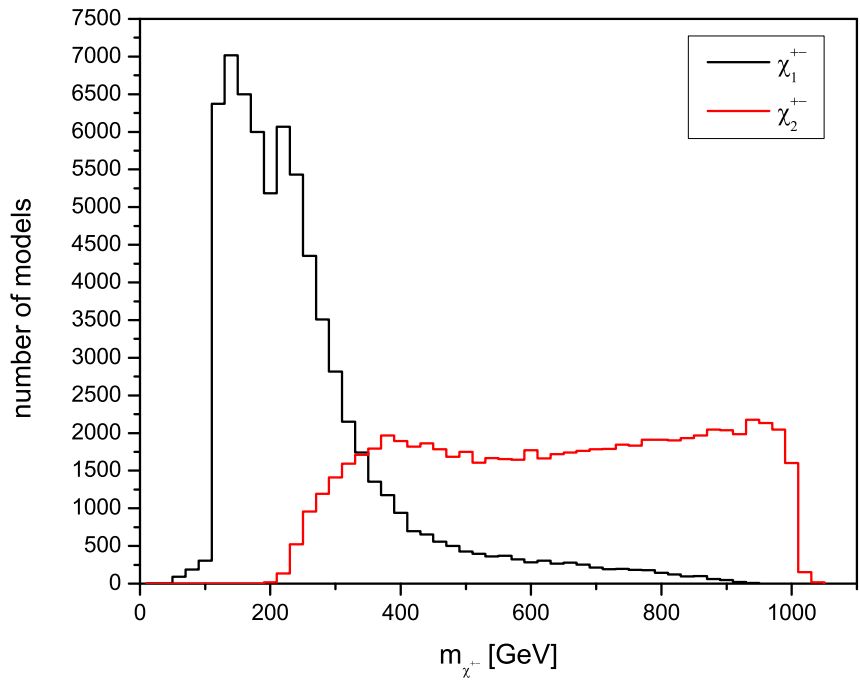}
\end{center}
\caption{Distribution of chargino masses for our set of
  models.}
\label{fig2}
\end{figure}

Figure~\ref{fig1} presents a histogram of the masses of the four
neutralino species in our models; Figure~\ref{fig2} displays a similar
histogram for the two chargino species.  The lightest neutralino is, of
course, the LSP.  The LSP mass lies between $100$ and $250$ GeV in over
$70\%$ of our models.  Generally models with a mostly Higgsino or Wino
LSP have a chargino with nearly the same mass as the LSP; as
sufficiently light
charginos would normally have been detected at LEP or the Tevatron, there
are fewer models with such LSPs with mass $\lsim 100$ GeV.

\begin{figure}[htbp]
\begin{center}
\includegraphics[width=14.0cm,angle=0]{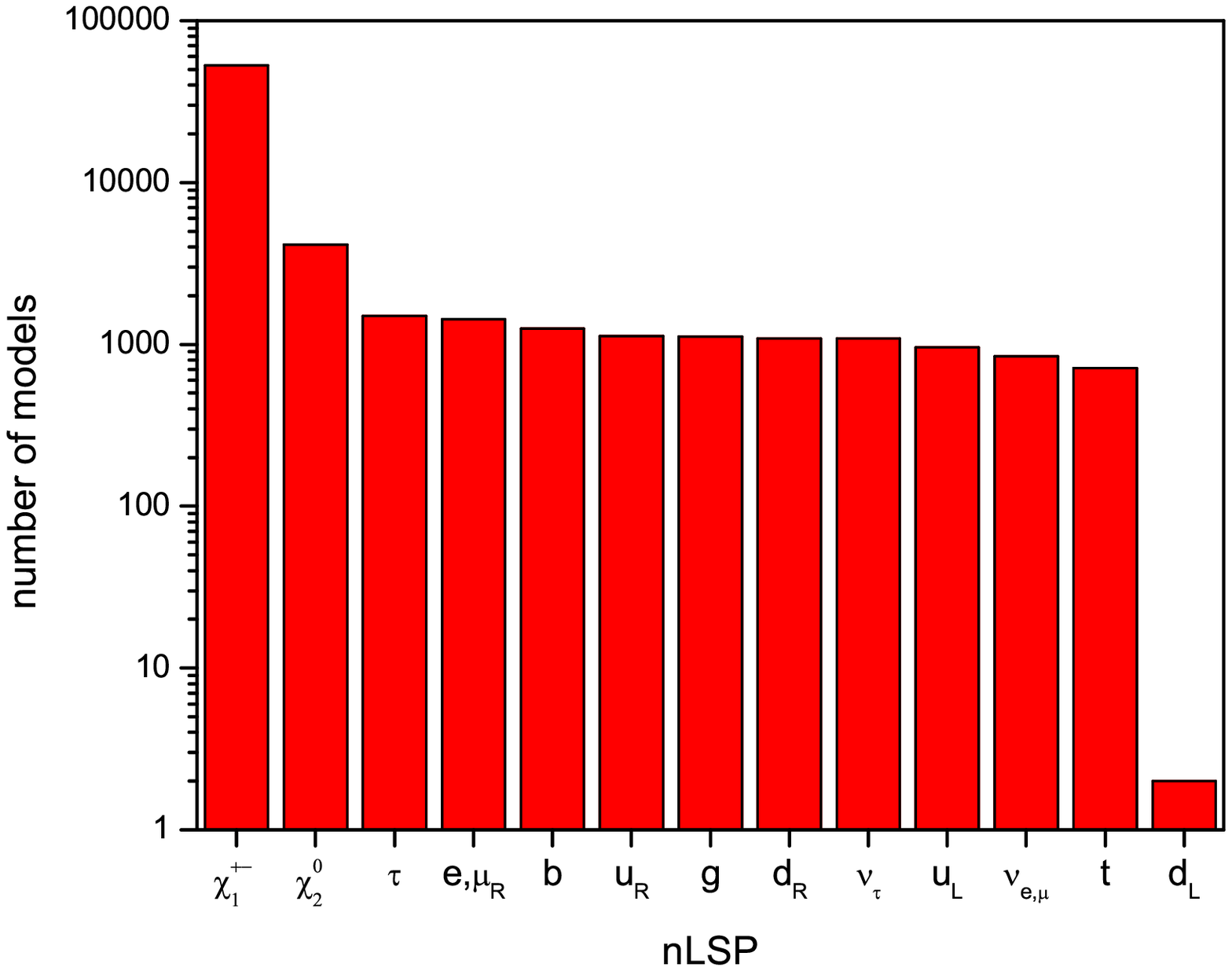}
\end{center}
\caption{Number of models in which the nLSP is the given sparticle.}
\label{fig3}
\end{figure}

The identity of the nLSP is shown in Figure~\ref{fig3}.  The lightest chargino
is the nLSP in about $78\%$ of the models; this is due to many 
models having Wino or Higgsino LSPs, and the generally small mass
splitting between a mostly Wino or Higgsino neutralino and the
corresponding chargino.  The second lightest
neutralino is the nLSP $\sim 6\%$ of the time.  These will generally
be models with a dominantly Higgsino LSP.  Note also that while
neutralinos or charginos are the nLSP in the vast majority of cases,
there are 10 other sparticles each of which is the nLSP in $>1\%$ of
our models.  Scenarios in which
these sparticles are the nLSP may lead to interesting
signatures at the LHC\cite{Us ATLAS}.

\begin{figure}[htbp]
\begin{center}
\includegraphics[width=14.0cm,angle=0]{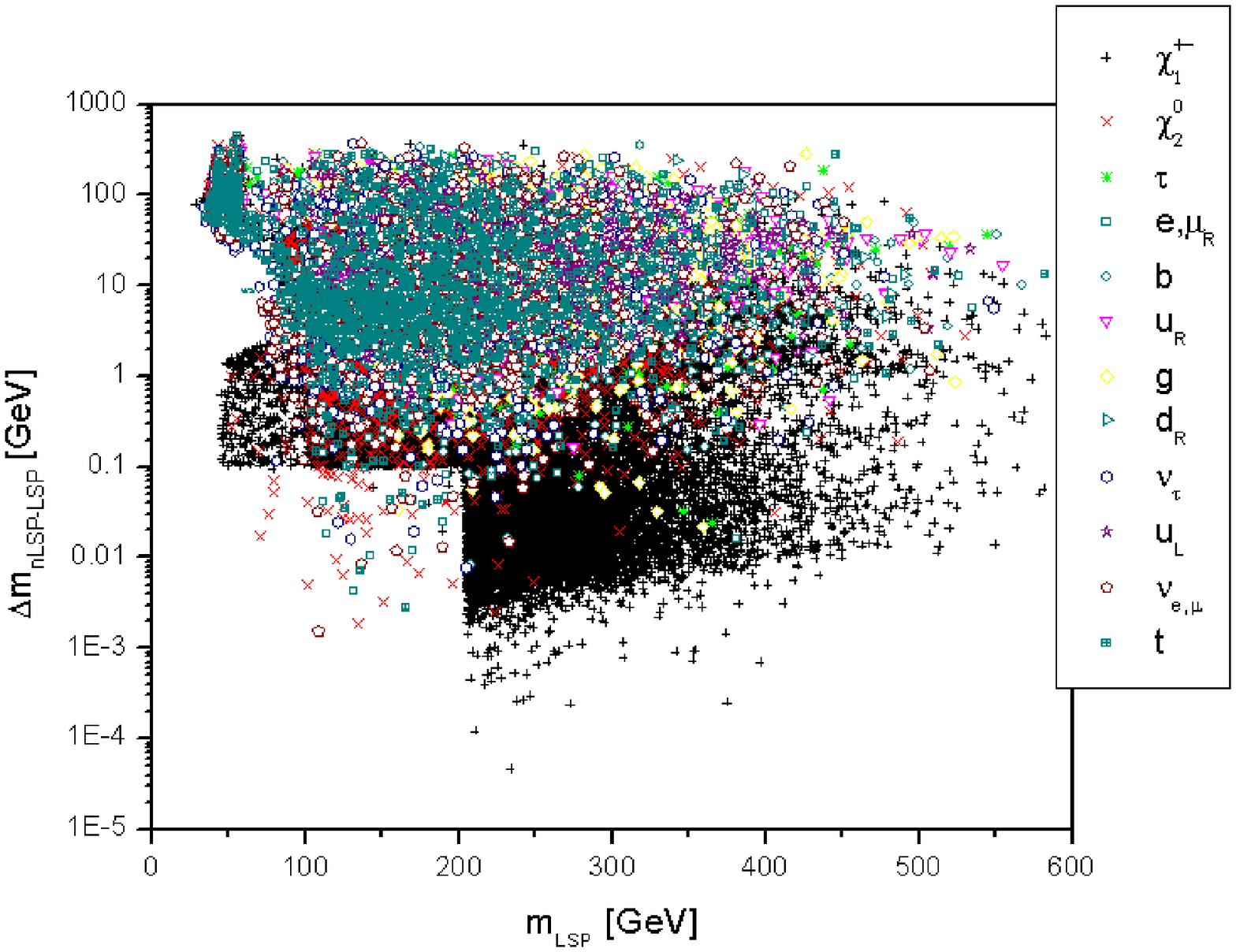}
\end{center}
\caption{Mass splitting between nLSP and LSP versus LSP mass.  The identity
of the nLSP is shown as well.  (The LSP is always the lightest neutralino in
our set of models).}
\label{fig4}
\end{figure}

Figure~\ref{fig4} displays the LSP mass value as a function of the 
LSP-nLSP mass
splitting, $\Delta m$,  our models for each identity of the LSP.  
It is interesting that these models have a smaller $\Delta m$ than
is often considered; $80\%$ of our models have $\Delta m < 10$ GeV,
$27\%$ have $\Delta m < 1$ GeV, and $3\%$ have $\Delta m < 10$ MeV.
As one can see from Figure~\ref{fig4}, this occurs largely, but not
exclusively, in models with a chargino nLSP. This is again due to
the many models where the LSP is nearly pure Wino or Higgsino.  

There are a number of interesting features in this figure.  The mostly
empty square region which appears on the lower left-hand side of
Figure~\ref{fig4} is due to the fact that models with chargino nLSPs
in this mass and $\Delta m$ range have been excluded by the Tevatron
stable chargino search.  Non-chargino nLSPs are not eliminated by this
search (\eg, the production cross section for sleptons in this range
is too small to be excluded by the Tevatron search).  It is perhaps
worth noting that a stable heavy charged particle search at the LHC,
corresponding to those done at the Tevatron, would be able to exclude
or discover the models with heavier chargino nLSPs and small values of
$\Delta m$ (corresponding to $\sim 12\%$ of our model set).

Another interesting feature in this figure is the bulge for $0.1$ GeV 
$\le \Delta m \lsim 2$ GeV and $m_{LSP}\lsim 100$ GeV.  
This region exists because
these values of $\Delta m$ are large enough that at LEP or the Tevatron,
the produced chargino would decay in the detector, but the resulting
charged tracks would be too soft to be observed.  The existence of such
a region shows the difficulty of making model independent statements
about sparticle masses or other SUSY observables.

We have seen that within our model set the nLSP can be almost any SUSY
particle and the corresponding $\Delta m$ can be small for these
cases.
Thus 
specific models in our set describe qualitatively most of the 
conventional long-lived sparticle scenarios.
Long-lived stops or staus (as in GMSB\cite{gmsbb}), gluinos (as in
Split SUSY\cite{SSS}) as well as charginos (as in AMSB\cite{amsbb})
all occur in our sample.  
We also have long-lived neutralinos, as does GMSB, however 
these are the $\tilde \chi_2^0$ in our case.
In addition to models which, to some extent, correspond to these
well-studied scenarios, we also have models with long-lived selectrons,
sneutrinos and sbottoms.

\begin{figure}[htbp]
\begin{center}
\includegraphics[width=14.0cm,angle=0]{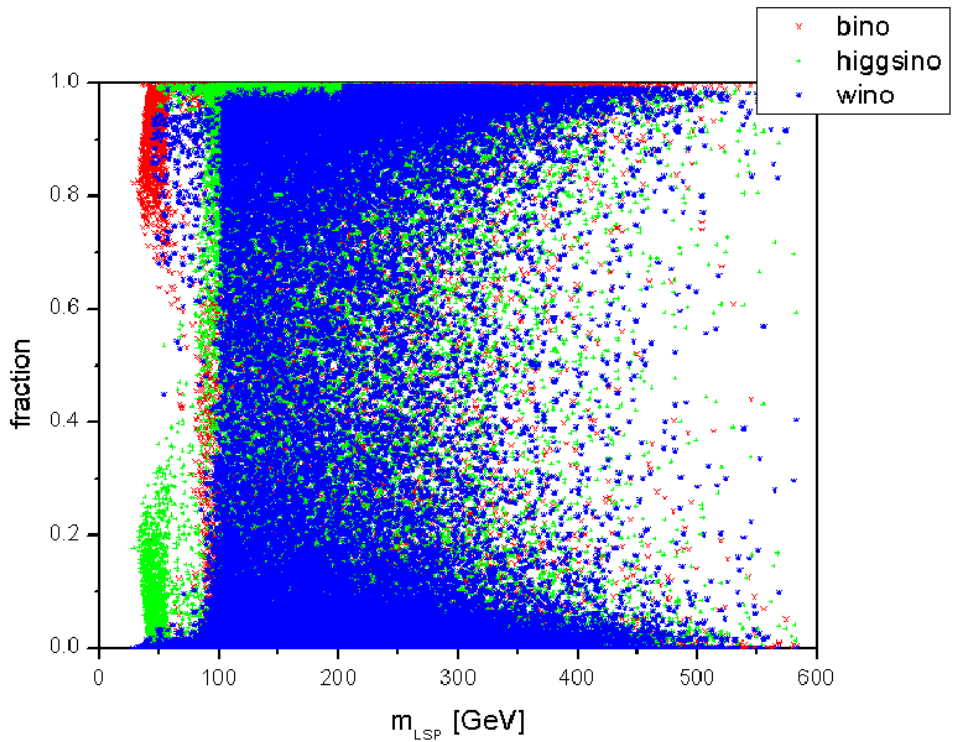}
\end{center}
\caption{The distribution of LSP gaugino eigenstate types as a function
  of the LSP mass.  Note that each LSP corresponds to three points on
  this figure, one each for its Bino, Wino, and Higgsino fraction.}
\label{fig5}
\end{figure}

\begin{figure}[htbp]
\begin{center}
\includegraphics[width=14.0cm,angle=0]{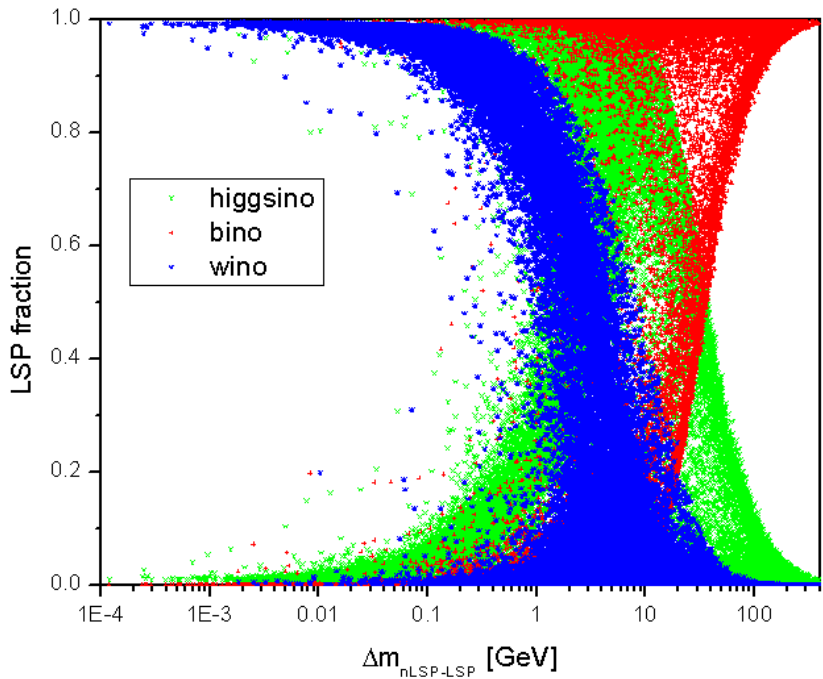}
\end{center}
\caption{The distribution of LSP gaugino eigenstate types as a function
  of the LSP-nLSP mass difference.  Note that as in Figure~\ref{fig5}, 
  every LSP corresponds to three points on this figure, one each for
  its Bino, Wino, and Higgsino fractions.}
\label{fig6}
\end{figure}
 
\begin{figure}[htbp]
\begin{center}
\includegraphics[width=14.0cm,angle=0]{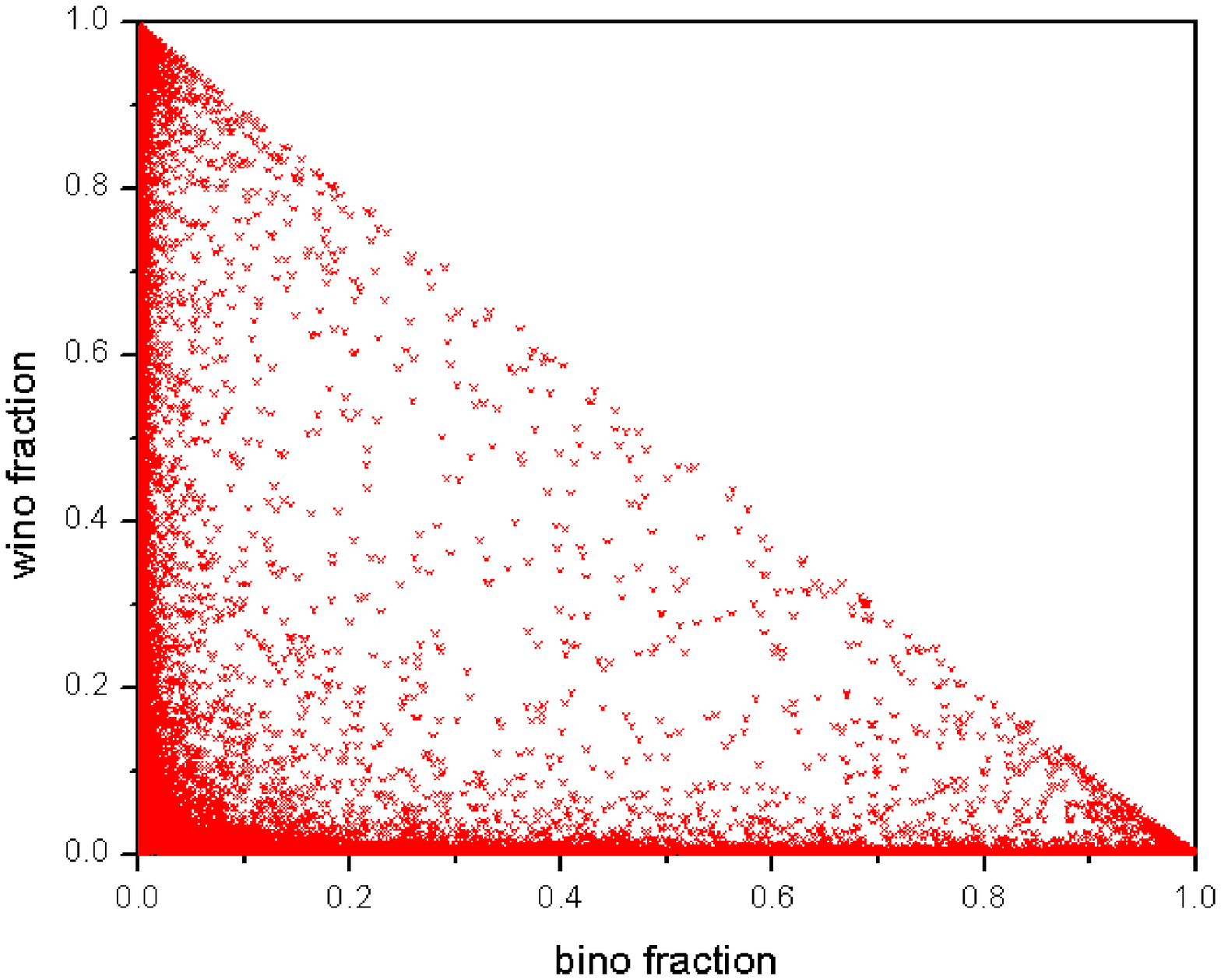}
\vspace*{0.1cm}
\includegraphics[width=14.0cm,angle=0]{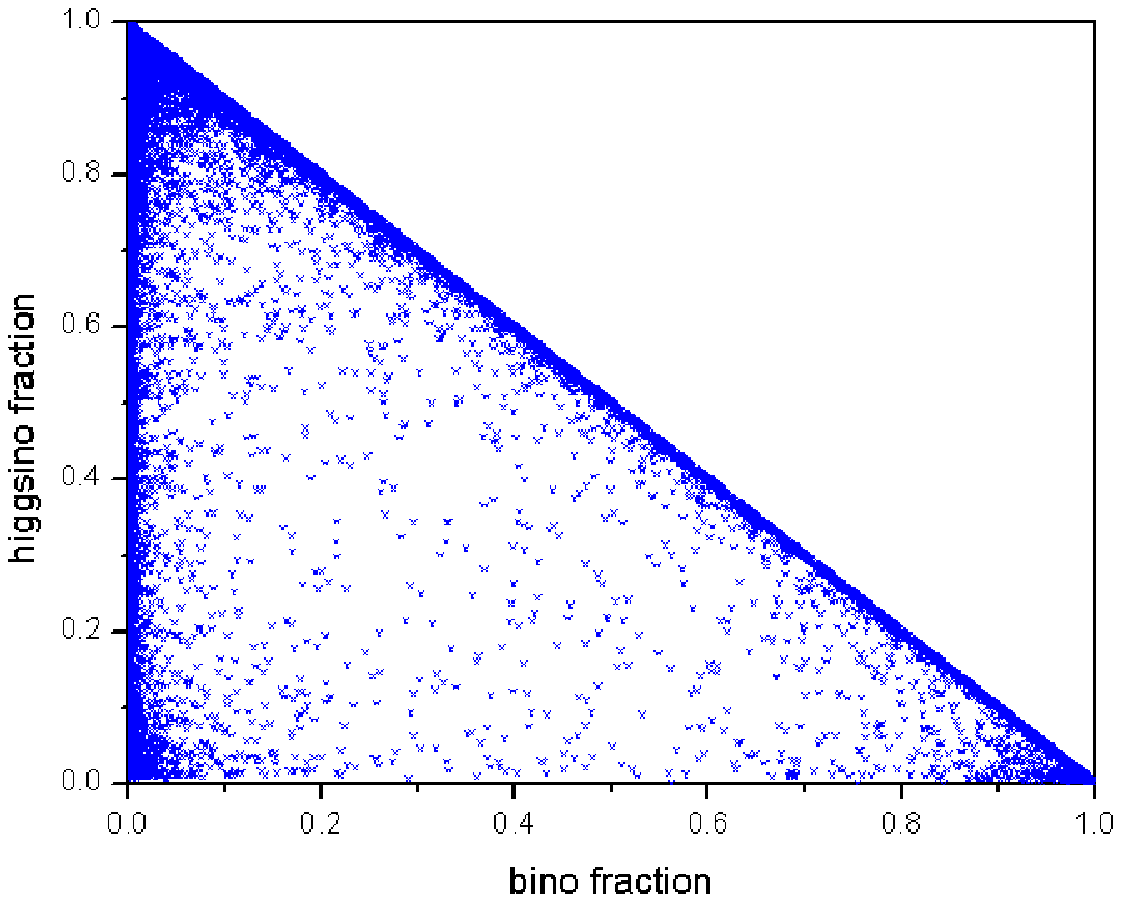}
\end{center}
\caption{Wino/Higgsino/Bino content of the LSP in the case of flat
  priors.  Note that, as elsewhere in the paper, $|Z_{11}|^2$,
  $|Z_{12}|^2$, and $|Z_{13}|^2 + |Z_{14}|^2$, where $Z_{ij}$ is the
  neutralino mixing matrix in the SLHA convention\cite{mssmrev}, give
  the Bino, Wino, and Higgsino fractions respectively.
 }
\label{fig7}
\end{figure}

Figures~\ref{fig5},~\ref{fig6}, and~\ref{fig7} display the gauge
eigenstate content of the LSPs in
our model set.  We note that most LSPs are relatively pure
eigenstates, with models where the LSP is Higgsino or mostly Higgsino
being by far the most common.  About one quarter of our models have
Wino or mostly Wino LSPs, while just over one-sixth have Bino or
mostly Bino LSPs. Within mSUGRA, the LSP is, in general, nearly purely
Bino; this suggests that most of our models are substantially
different from mSUGRA.   A more precise breakdown of the
content of LSPs in the model set is presented in Table~\ref{neutralino
  mixing}.  
We note that one would expect the LSP be a pure eigenstate
fairly often in a random scan of Lagrangian parameters, since if the
differences between $M_1,M_2,$ and $\mu$ are large compared to $M_Z$,
then the eigenstates of the mixing matrix will be essentially pure
gaugino and Higgsino states\cite{mssmrev}.  

\begin{table}
\centering
\begin{tabular}{|l|c|r|} \hline\hline 
LSP Type & Definition & Fraction \\ 
& & of Models \\ \hline
Bino & $|Z_{11}|^2 > 0.95$ & 0.14 \\
Mostly Bino & $0.8 < |Z_{11}|^2 \leq 0.95$ & 0.03 \\
Wino & $|Z_{12}|^2 > 0.95$ & 0.14 \\
Mostly Wino & $0.8 < |Z_{12}|^2 \leq 0.95$ & 0.09\\
Higgsino & $|Z_{13}|^2+|Z_{14}|^2 > 0.95$ & 0.32 \\
Mostly Higgsino & $0.8 < |Z_{13}|^2+|Z_{14}|^2 \leq 0.95$ &  0.12 \\
All other models & & 0.15 \\
\hline\hline
\end{tabular}
\caption{The fractions of our model set for which the LSP is of each
  of the given types.  These types are defined here by the modulus squared
  of elements of neutralino mixing matrix in the SLHA convention.
  See~\cite{mssmrev} for details.}
\label{neutralino mixing}
\end{table}

\subsection{Relic Density}

\begin{figure}[htbp]
\begin{center}
\includegraphics[width=9.0cm,angle=-90]{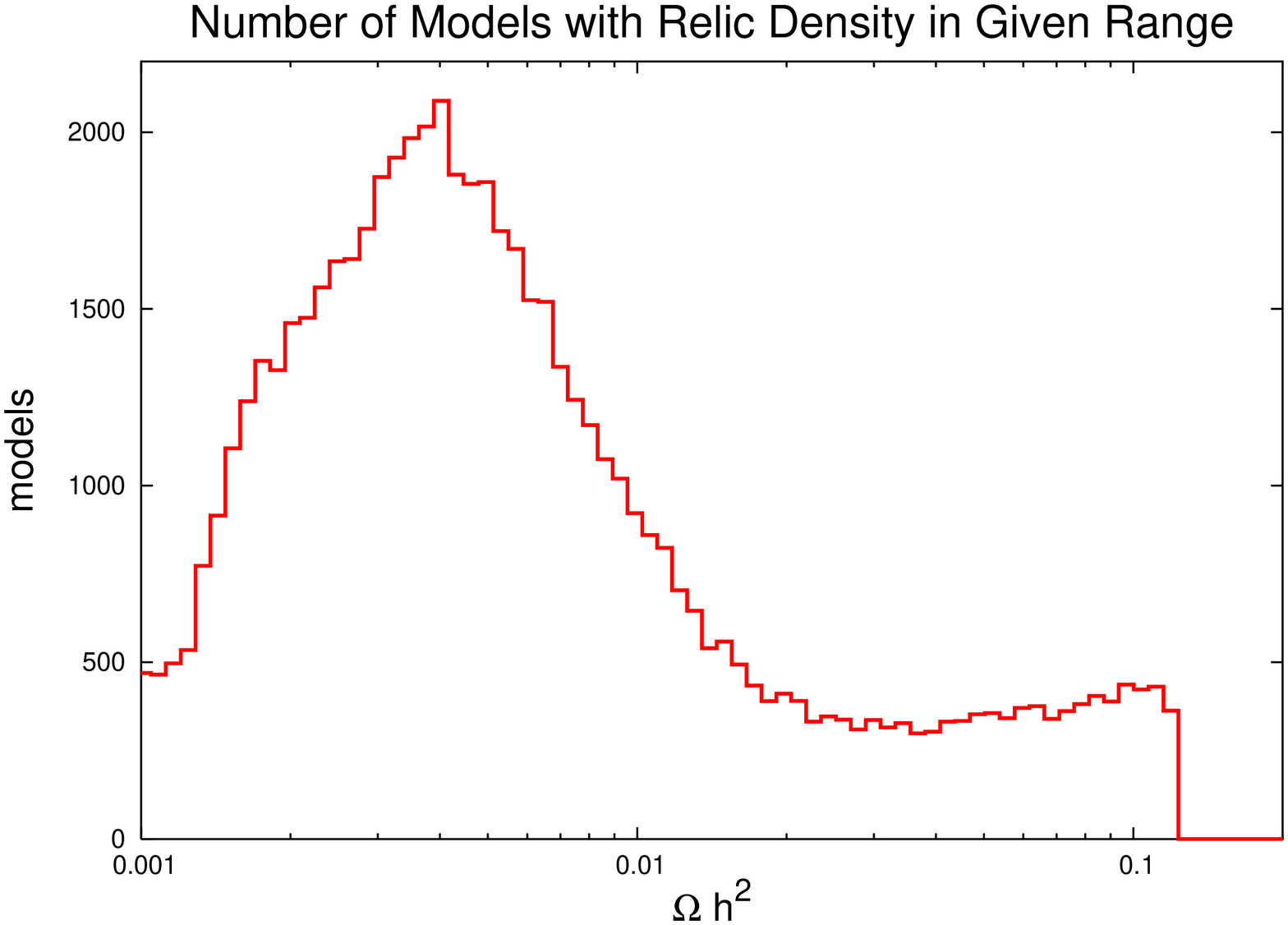}
\end{center}
\caption{Distribution of $\Omega
h^2|_{\mathrm{LSP}}$ for our models.}
\label{fig8}
\end{figure}

We did not demand that the LSP, in any given model, account for
all of the dark matter, rather we required only that the LSP relic
density not be too large to be consistent with WMAP.  More
specifically, we employed $\Omega h^2|_{\mathrm{LSP}} < 0.121$.  
Figure~\ref{fig8} shows the
distribution of $\Omega h^2|_{\mathrm{LSP}}$ values predicted by our model
set.  Note that this distribution is peaked at small values of $\Omega
h^2|_{\mathrm{LSP}}$.  In particular, the mean value for this
quantity in our models is $\sim 0.012$, which is about ten times less than
the central value determined from the aforementioned WMAP and supernova
data\cite{Komatsu:2008hk}. 
We note that the range of possible values of $\Omega
h^2|_{\mathrm{LSP}}$ is found to be much larger than those obtained by
analyses of specific SUSY breaking scenarios{\cite {big}}.

\begin{figure}[htbp]
\begin{center}
\includegraphics[width=9.0cm,angle=-90]{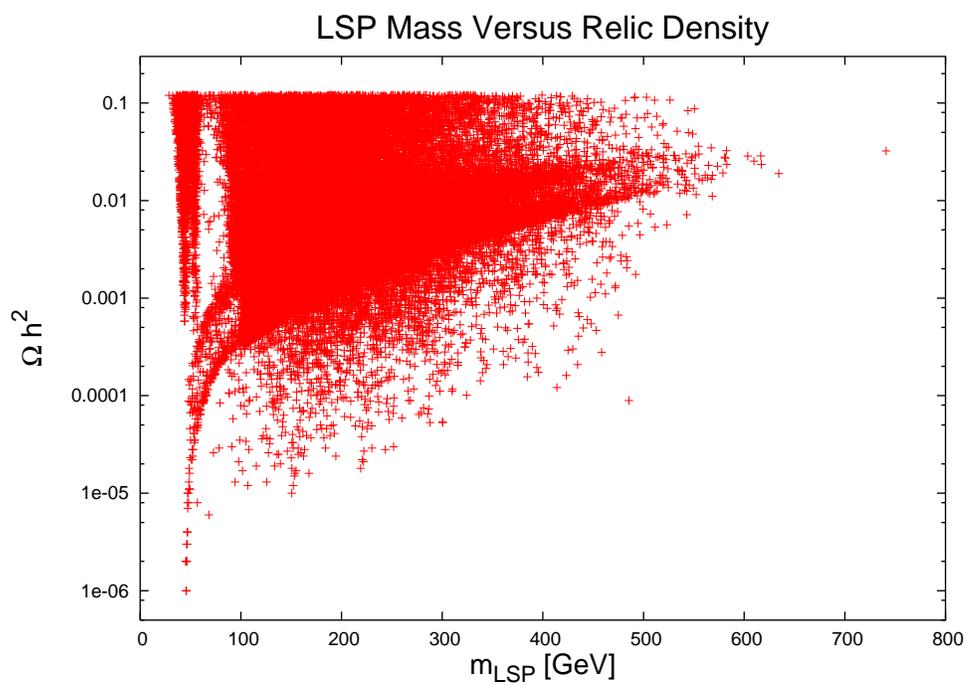}
\end{center}
\caption{Distribution of $\Omega
h^2|_{\mathrm{LSP}}$ as a function of
  the LSP mass.}
\label{fig9}
\end{figure}

\begin{figure}[htbp]
\begin{center}
\includegraphics[width=9.0cm,angle=-90]{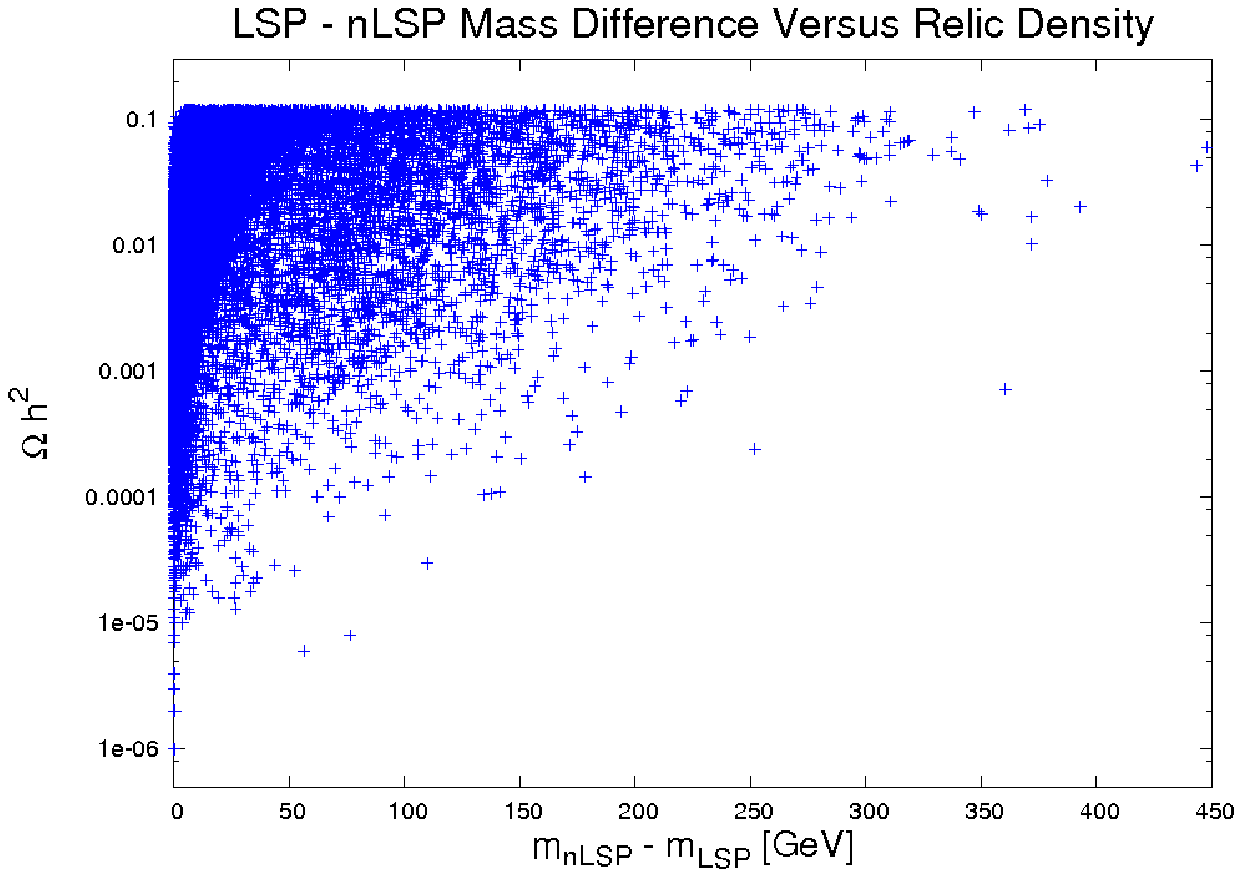}
\end{center}
\vspace{-.2 in}
\caption{Distribution of $\Omega
h^2|_{\mathrm{LSP}}$ as a function of
  the LSP-nLSP mass splitting.}
\label{fig10}
\end{figure}

We display the predictions for $\Omega
h^2|_{\mathrm{LSP}}$ versus the LSP mass in
Figure~\ref{fig9} and versus the nLSP - LSP mass splitting in
Figure~\ref{fig10}.  Figure~\ref{fig9} makes it clear that
$\Omega h^2|_{\mathrm{LSP}}$ generally increases 
with the LSP mass, but a large
range of values for the relic density are possible at any given LSP mass.  
The empty region in Figure~\ref{fig9} where $\Omega h^2|_{\mathrm{LSP}} \approx 0.001 -
0.1$ and $m_{\mathrm{LSP}}\approx 50-100$ is due to the fact that, in
general, LSPs which are mostly Higgsino or Wino give lower values of
$\Omega h^2|_{\mathrm{LSP}}$, and, as noted above, there are fewer
Higgsino or Wino LSPs in this mass range.
Figure~\ref{fig10} shows that small mass differences can lead to large
dark matter annihilation rates.

\subsection{Direct Detection of Dark Matter}

As noted above, we calculate the spin-dependent and spin-independent
WIMP-nucleon cross sections using micrOMEGAs 2.21~\cite{MICROMEGAS}.
These data give the possible signatures in our
model set for experiments that search for WIMPs directly.  As these
experiments measure the product of WIMP-nucleon cross
sections with the local relic density, the cross section data presented in
the figures below are scaled by $\xi = \Omega
h^2|_{\mathrm{LSP}} / \Omega h^2|_{\mathrm{WMAP}}$.  To date,
  these experiments generally provide a more significant bound on the
  spin-independent cross section, and hence we will focus on those.

\begin{figure}[htbp]
\begin{center}
\includegraphics[width=9.0cm,angle=-90]{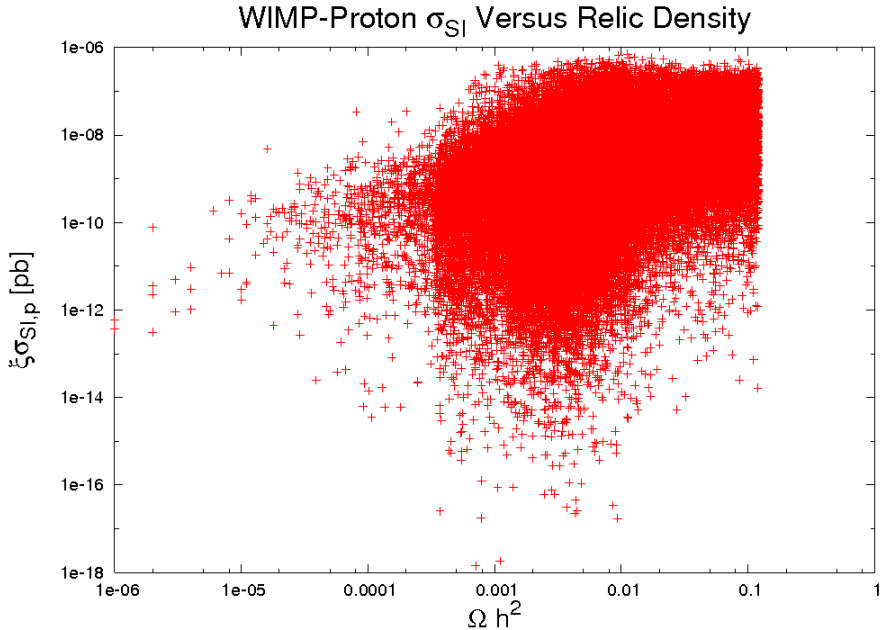}
\end{center}
\caption{
Distribution of scaled WIMP-proton spin-independent cross section
versus the LSP contribution to relic density for our models.
}
\label{fig11}
\end{figure}

Figure~\ref{fig11} presents the distribution for the scaled WIMP-proton
spin-independent cross section versus relic density for our model
sample.  As one would expect, larger values of the cross section
 are generally found at larger values of $\Omega h^2|_{\mathrm{LSP}}$.  
However, even for relic densities close to the
  WMAP value, $\xi \sigma_{p,SI}$ is seen to vary by almost eight
  orders of magnitude.
These ranges for $\xi \sigma_{p,SI}$ are much larger than those from
mSUGRA as calculated, \eg, in~\cite{Barger:2008qd}. 

\begin{figure}[htbp]
\begin{center}
\includegraphics[width=9.0cm,angle=-90]{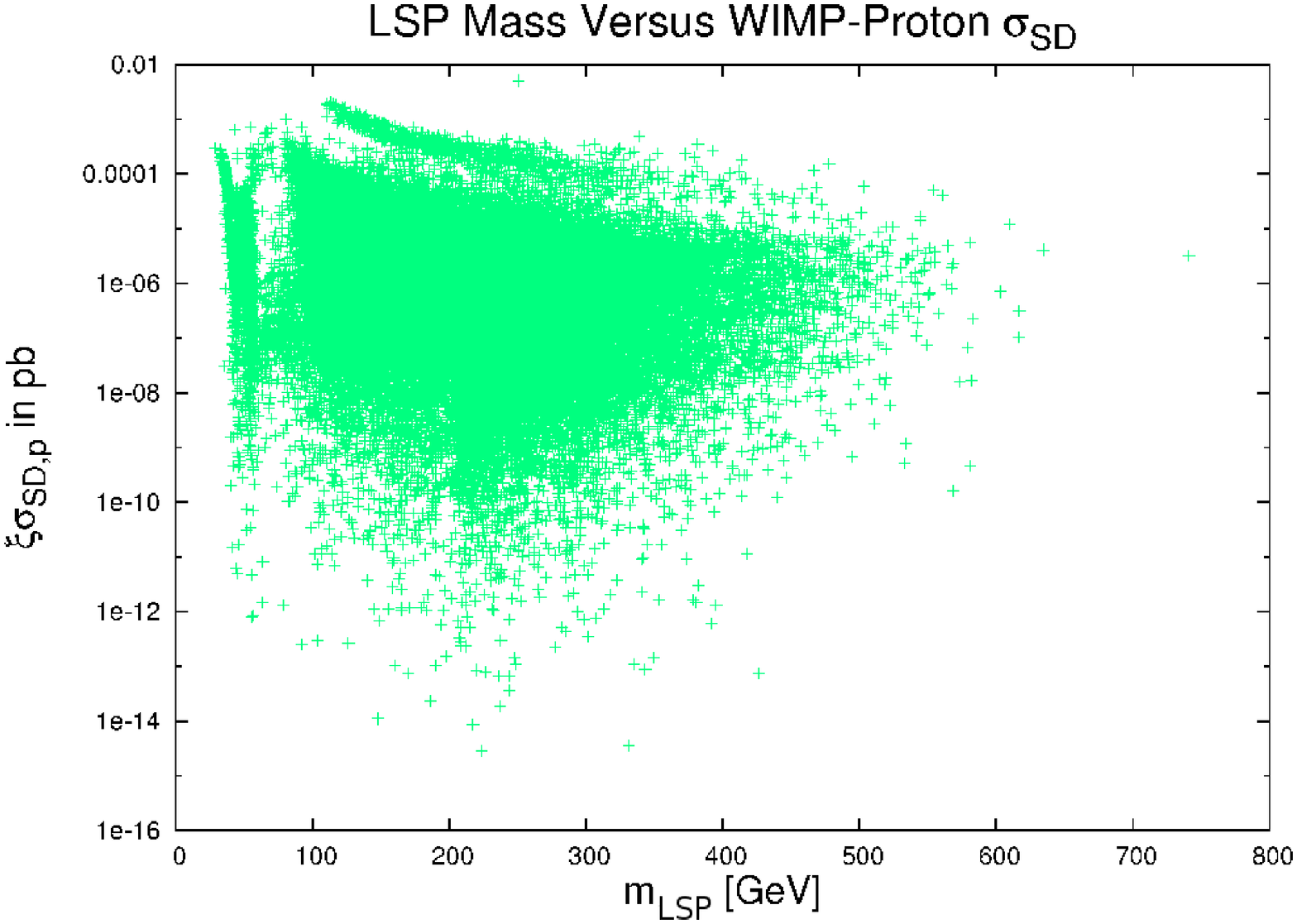}
\vspace*{1cm}
\includegraphics[width=9.0cm,angle=-90]{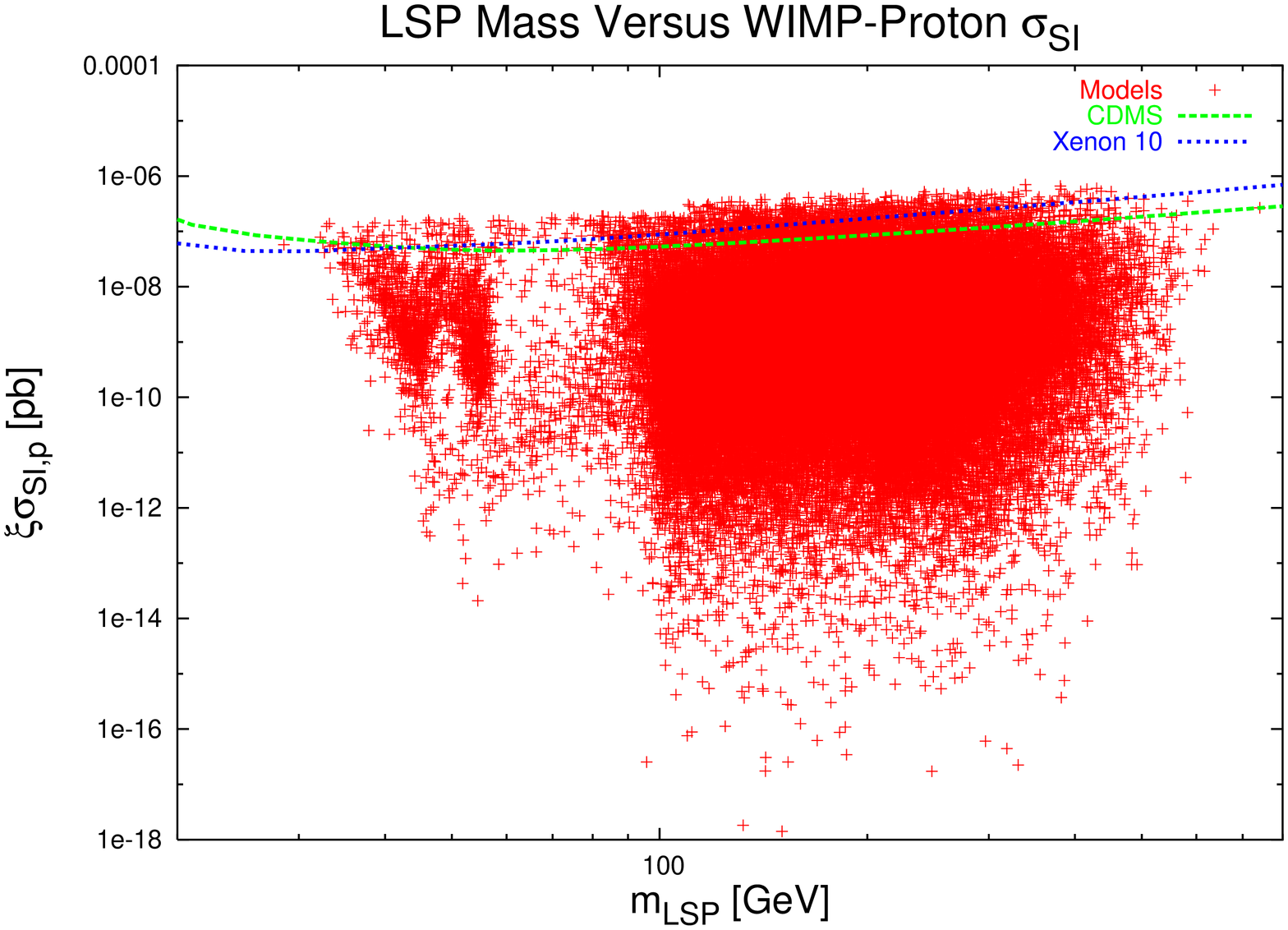}
\end{center}
\caption{
Distributions of scaled WIMP-proton spin-dependent cross section and
spin-independent cross sections versus LSP mass in our models.  In the
spin-independent panel, the CDMS and Xenon10 bounds, which provide
the strongest limits for the range in LSP mass relevant for our
models, are shown.
 }
\label{fig12}
\end{figure}

Figure~\ref{fig12} shows the scaled WIMP-proton spin-dependent and
spin-independent cross sections as a function of the LSP mass.  The
constraints from XENON10\cite{XENON10} and CDMS\cite{CDMS} are also
displayed.  
As noted above, to take the uncertainties in the theoretical
calculations of the WIMP-nucleon cross section into account, we
allowed for a factor of 4 uncertainty in the calculation of the WIMP-nucleon
cross section.
Table~\ref{exclude} gives the
fraction of models that would be excluded if the combined
CDMS/XENON10 cross section limit
were improved by an overall scaling factor. 
Note that our inclusion of the theoretical uncertainties does not
significantly modify the size of our model sample.
\begin{table}
\centering
\begin{tabular}{|c|c|} \hline\hline
Improvement in S.I. & Fraction of Models \\
Cross Section Limit & Excluded \\ \hline
4     &   0.032 \\
10    &   0.071 \\
40    &   0.19 \\
100   &   0.31 \\
400   &   0.52 \\
1000  &   0.65 \\
4000  &   0.81 \\
\hline\hline
\end{tabular}
\caption{The fraction of our model set which would be excluded
  for the specified improvement in the direct detection bound on the
  spin-independent WIMP-nucleon cross section.}
\label{exclude}
\end{table}

We find that the range of values obtained for these cross sections
covers the entire
region in cross section/ LSP space that is anticipated from different
types of Beyond the Standard Model theories in the above reference.
This possibly suggests that we cannot use direct detection experiments to
distinguish between \eg~SUSY versus Little Higgs versus Universal
Extra Dimensions dark matter candidates in the absence of other data.

\begin{figure}[htbp]
\begin{center}
\includegraphics[width=9.0cm,angle=-90]{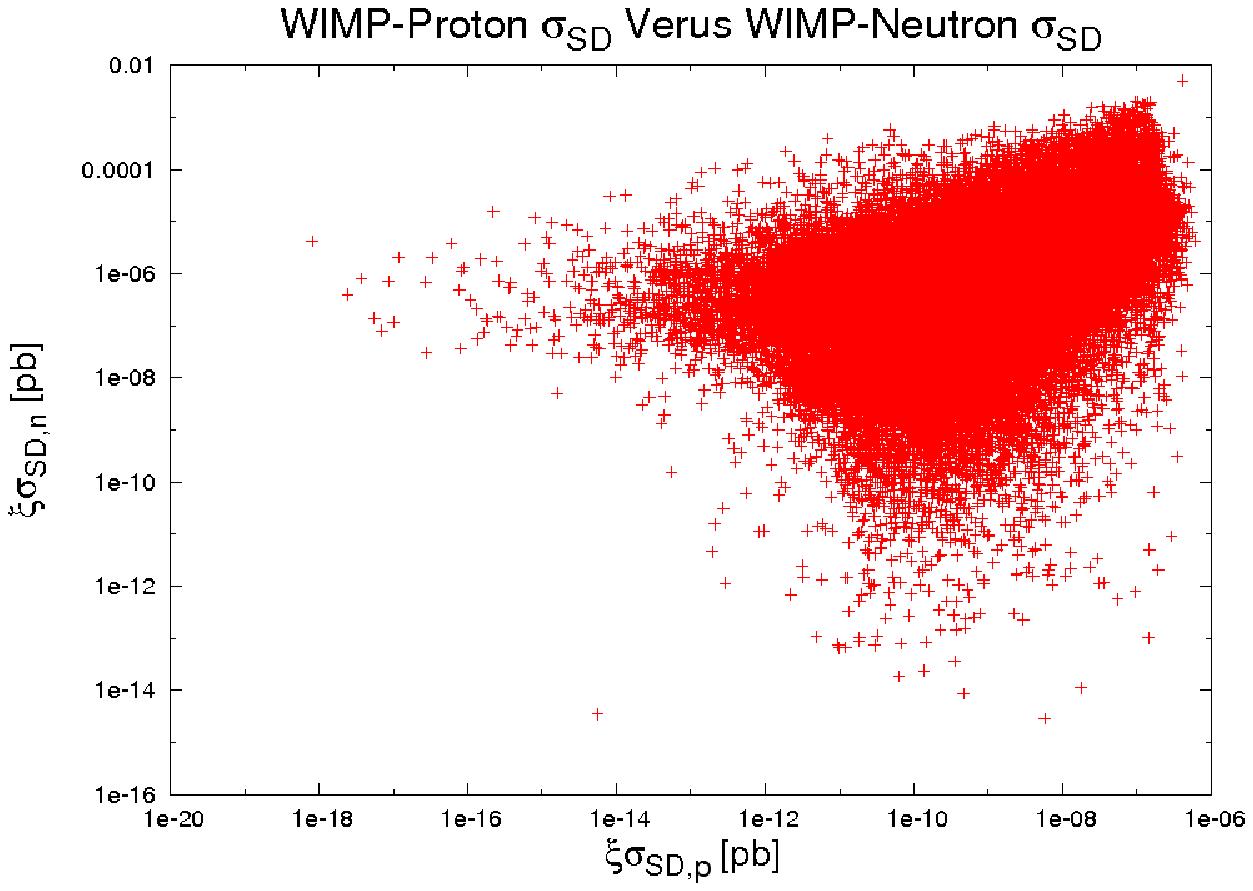}
\vspace*{0.1cm}
\includegraphics[width=9.0cm,angle=-90]{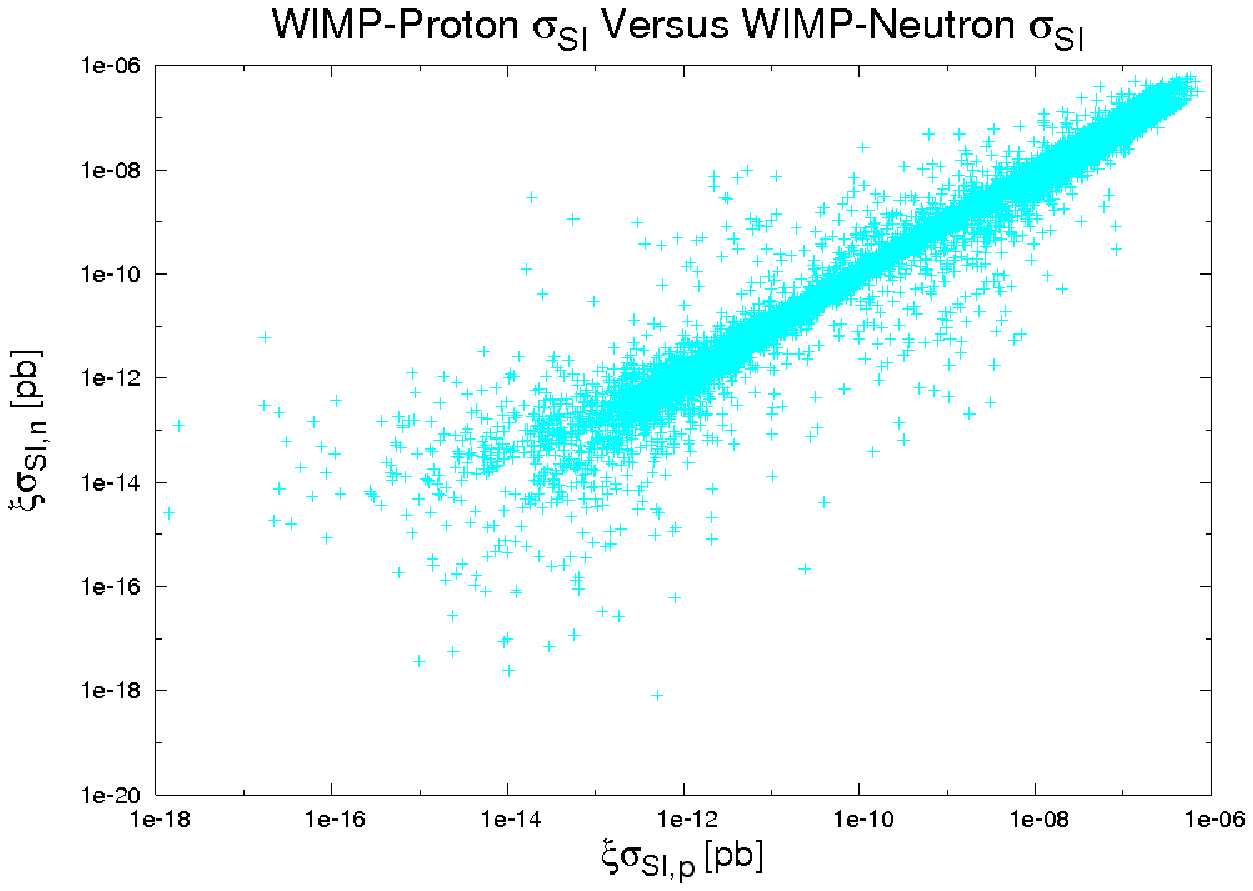}
\end{center}
\caption{
Here we compare WIMP-neutron and WIMP-proton cross sections.  The
spin-dependent cross sections are shown in the top panel; the
spin-independent cross sections in the bottom panel.
 }
\label{fig13}
\end{figure}

In Figure~\ref{fig13}, we compare the WIMP-proton and WIMP-neutron cross
sections in the spin-dependent and spin-independent cases.  The
spin-independent cross sections are seen to be fairly isospin independent;
this is not the case, however, for the spin-dependent cross sections.

\subsection{Indirect Detection of Dark Matter}

The PAMELA collaboration has recently claimed an excess in the ratio
of cosmic ray positrons to electrons observed at energies
$\gsim 10$ GeV\cite{Adriani:2008zr}.
Here we employ DarkSUSY 5.0.4\cite{DarkSUSY} to calculate this ratio for
our model sample and compare these results with the PAMELA data.

In general, for a thermal relic dark matter candidate to reproduce
the PAMELA data, its signal rate must be multiplied by a boost 
factor\cite{boost}.  
In nature, such a boost factor could
result from, \eg, a local overdensity.  The boost factor in that case
would be the square of the ratio between the density of dark matter in
the region from which one is sensitive to cosmic ray positrons and
electrons to the universe as a whole.

We have investigated four of the propagation models available as default choices in DarkSUSY: the
model of Baltz and Edsj\"{o}(BE)\cite{Baltz:1998xv}, that of Kamionkowski
and Turner(KT)\cite{Kamionkowski:1990ty}, that of Moskalenko and
Strong(MS)\cite{Moskalenko:1999sb}, as well as
GALPROP\cite{GALPROP}.  In the figures that follow we show the results of calculations using the MS 
propagation model, which is based on early GALPROP Green's functions and whose results typically lie between those computed otherwise.
However, we note that the extent to which the 
positron/electron flux ratio predicted by our models matches the PAMELA
data can be highly sensitive to the choice of propagation model
parameters and assumed astrophysical backgrounds.  We will explore this further in future 
work\cite{Us  DM}. The halo model employed here is the Navarro-Frenck-White
profile\cite{NFW}.

\begin{figure}[htbp]
\begin{center}
\includegraphics[width=9.0cm,angle=-90]{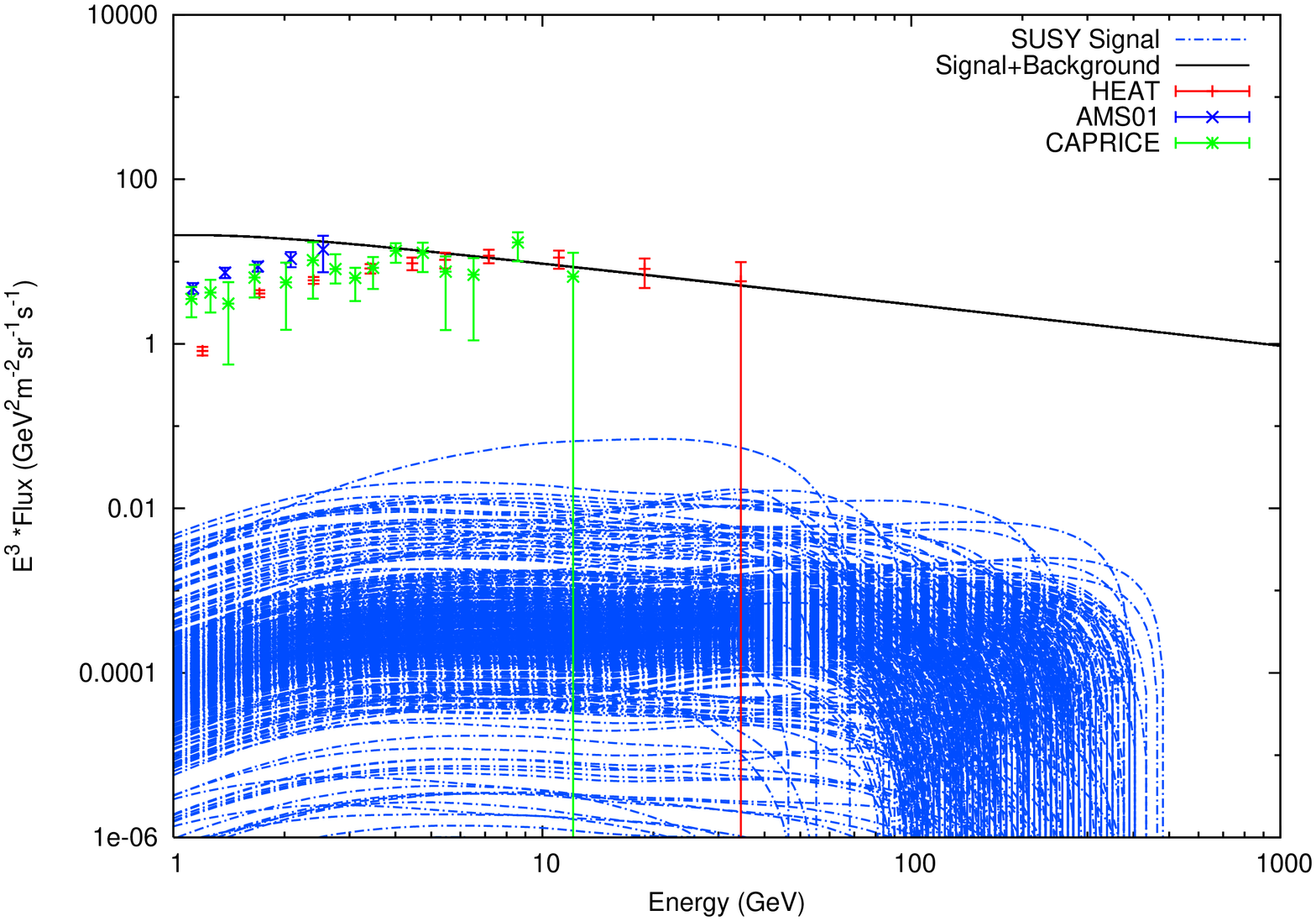}
\end{center}
\caption{
Expected, $E^3$ scaled, contribution to the absolute flux of positrons (unboosted) from neutralino annihilation in
the halo for 500 randomly selected models employing MS propagation. Also shown are data from HEAT\cite{DuVernois:2001bb}, AMS01\cite{Aguilar:2002ad}, and CAPRICE94\cite{CAPRICE} as well as  
the DarkSUSY default secondary positron background (a parameterization of model 08-005 in\cite{Moskalenko:1997gh}.}
\label{fig14}
\end{figure}

\begin{figure}[htbp]
\begin{center}
\includegraphics[width=9.0cm,angle=-90]{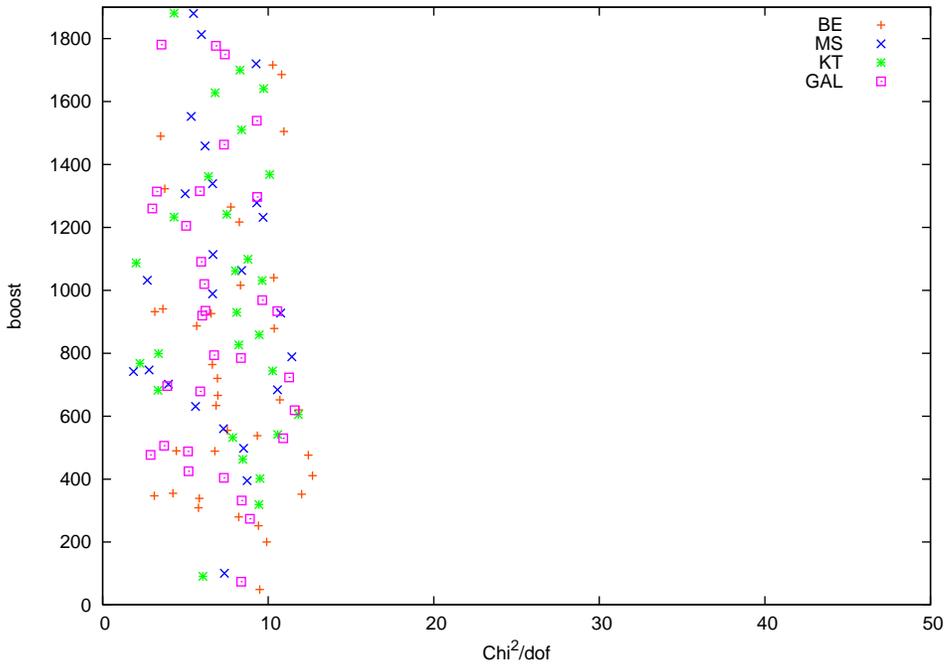}
\end{center}
\caption{
The distribution of $\chi^2$ per degree of freedom versus the choice
of boost factor that minimized this quantity for 500 randomly selected
pMSSM models in our model set for the four propagation models discussed in the text.}
\label{fig15}
\end{figure}

\begin{figure}[htbp]
\begin{center}
\includegraphics[width=9.0cm,angle=-90]{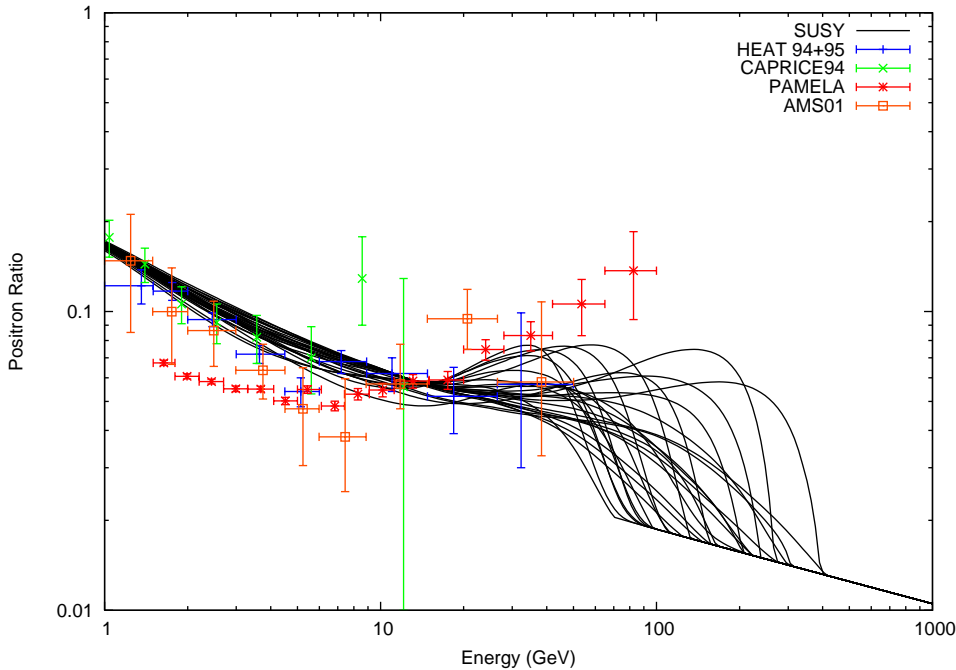}
\end{center}
\caption{
Positron/ electron flux ratio versus energy taking models from a set of 500 random pMSSM models
for which the $\chi^2$ per degree of freedom with the
$\chi^2$-maximizing boost was less than 10.0 for three of the four propagation models studied (curves shown use MS propagation). Data shown are from HEAT\cite{Barwick:1997ig}, CAPRICE94\cite{CAPRICE}, PAMELA\cite{Adriani:2008zr} and AMS01\cite{Aguilar:2002ad}.}
\label{fig16}
\end{figure}

The differential positron flux as a function of energy for 
a random sample of 500 models from our set are shown  in
Figure~\ref{fig14}.  Here we assume a boost factor of $1$; the
normalization of the curves takes into account the fact that for many
of these models  $\Omega h^2|_{\mathrm{LSP}} < \Omega_{\mathrm{WMAP}}$.
  
We next determine how well the predicted positron fluxes for these
models agree with the PAMELA data, allowing for the possibility of a
boost factor.  To do this, we find the value for the boost factor
(with the restriction that it be $< 2000$)
which minimizes the $\chi^2$
for the fit of each model's prediction to the PAMELA data. (Note that many of the 
models require an even larger boost to obtain a good fit and are thus not 
shown in Figure~\ref{fig15}). In
calculating the $\chi^2$, we consider only the seven highest energy
bins, as at lower energies solar modulation is expected to play a major
role\cite{Adriani:2008zr}.  Figure~\ref{fig15} shows the $\chi^2$ and
corresponding boost factor for these 500 random models.
Note that there are four data points for each model in this figure.  
We then display the positron to electron flux ratio, for the models
with a low value of $\chi^2$ employing MS propagation, as a function of energy in 
Figure~\ref{fig16}, and note the reasonable agreement with the data
for some models.

\begin{figure}[htbp]
\begin{center}
\includegraphics[width=9.0cm,angle=-90]{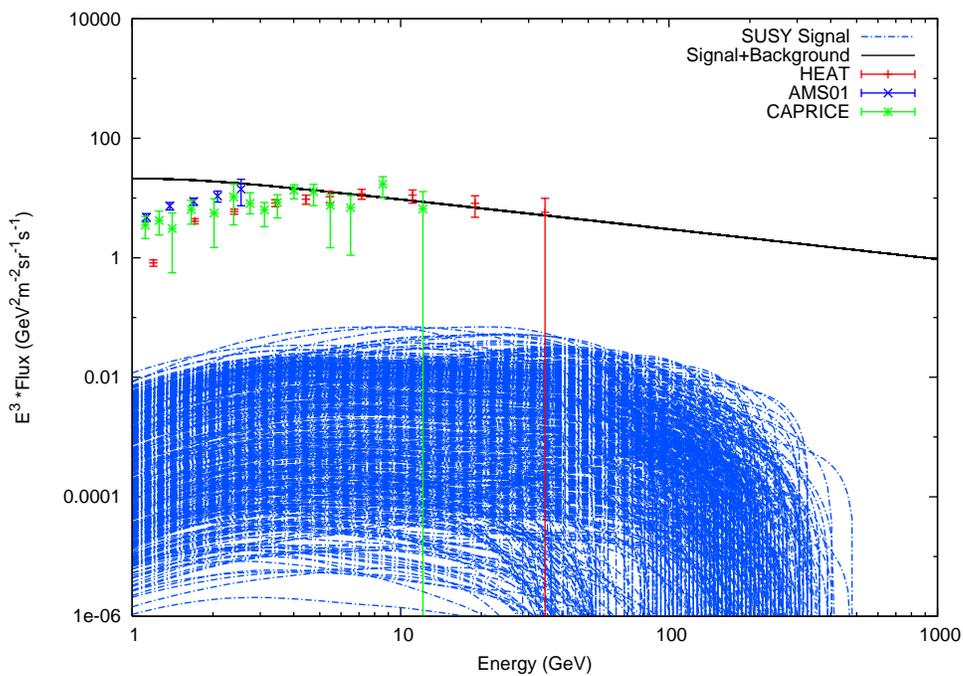}
\end{center}
\caption{Same as Fig.~14 but now for a 500 model set satisfying 
$\Omega h^2|_\mathrm{WMAP} \ge \Omega h^2|_\mathrm{LSP}>0.10$.}
\label{fig17}
\end{figure}

\begin{figure}[htbp]
\begin{center}
\includegraphics[width=9.0cm,angle=-90]{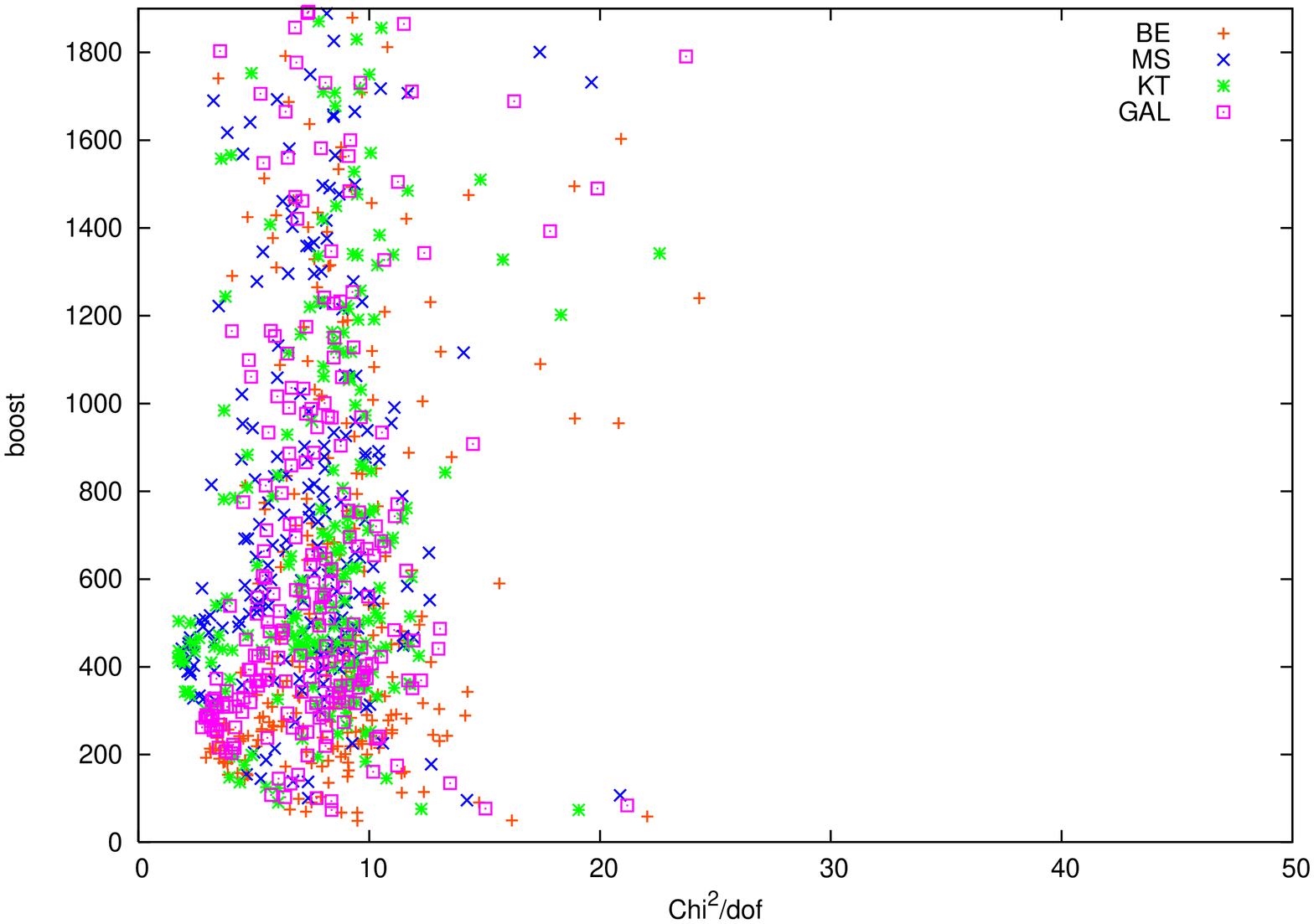}
\end{center}
\caption{Same as Fig.~15 but now for a 500 model set satisfying 
$\Omega h^2|_\mathrm{WMAP} \ge \Omega h^2|_\mathrm{LSP}>0.10$.}
\label{fig18}
\end{figure}

\begin{figure}[htbp]
\begin{center}
\includegraphics[width=9.0cm,angle=-90]{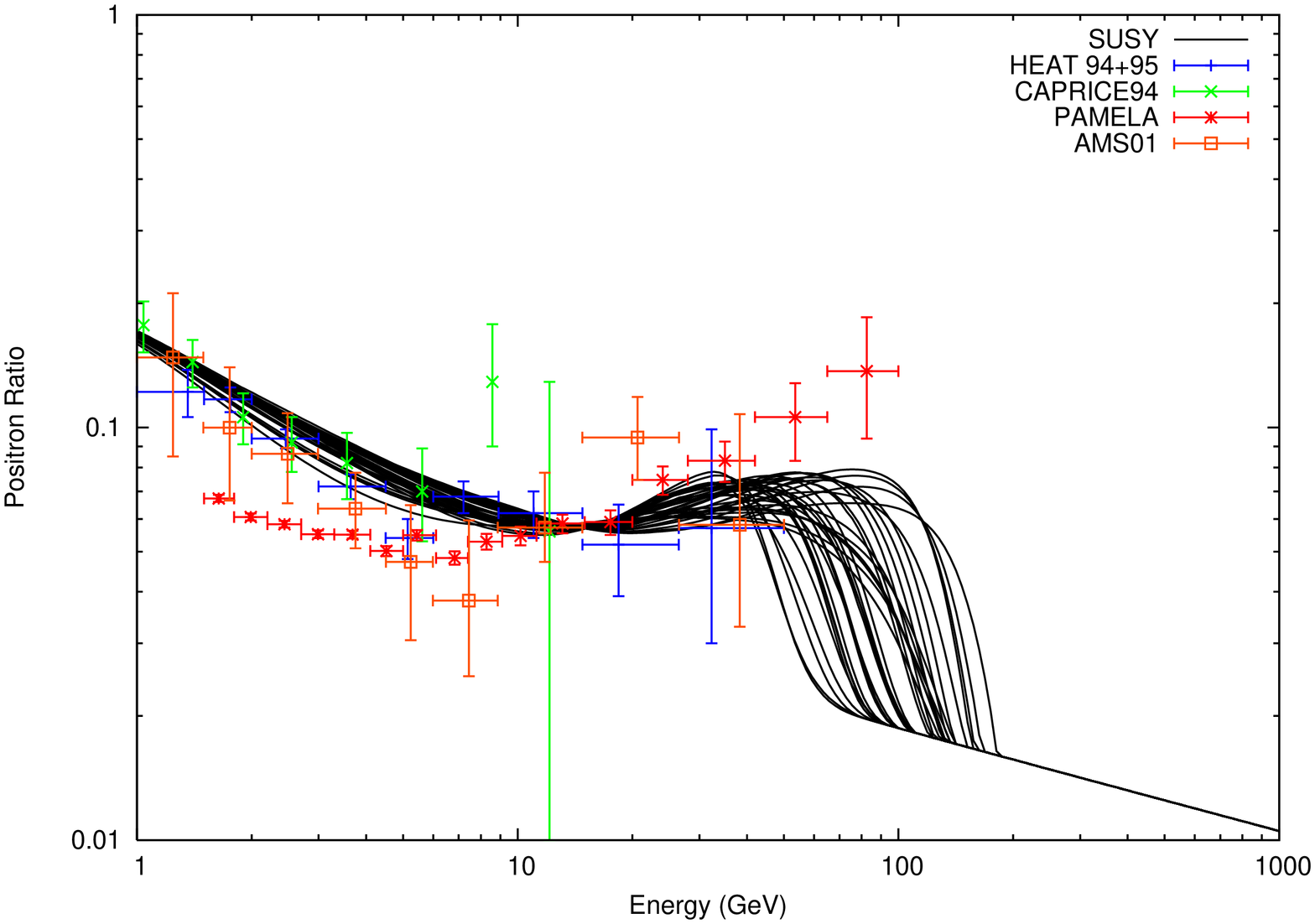}
\end{center}
\caption{Positron/ electron flux ratio versus energy taking models from a set of 500 pMSSM models satisfying $\Omega h^2|_\mathrm{WMAP} \ge \Omega h^2|_\mathrm{LSP}>0.10$ and for which $\chi^2$-maximizing boost was less than 5.0 for three of the four propagation models studied (curves shown use MS propagation). Data shown are from HEAT\cite{Barwick:1997ig}, CAPRICE94\cite{CAPRICE}, PAMELA\cite{Adriani:2008zr} and AMS01\cite{Aguilar:2002ad}.}
\label{fig19}
\end{figure}

Since the flux from WIMP annihilation scales as $(\Omega
h^2|_{\mathrm{LSP}}/\Omega h^2|_{\mathrm{WMAP}})^2$, we might expect to
improve the match to the PAMELA data using models from our sample
for which
$\Omega h^2|_{\mathrm{LSP}}\approx \Omega h^2|_{\mathrm{WMAP}}$.  To
test this, we examine the predicted positron flux for 500 random models
with $\Omega h^2|_{\mathrm{LSP}} > 0.100$ employing MS propagation;
these fluxes are shown in Figure~\ref{fig17} with no boost factor.  
We then again find the boost 
factor that minimizes the $\chi^2$ of the positron to electron flux
ratios with respect to the seven highest energy PAMELA bins;
these are shown in Figure~\ref{fig18}.  Here, we note that there are many more
models for which the $\chi^2$-minimizing value for the boost factor is $<
2000$ and there are many more points for which the $\chi^2$ value is low.  
The positron to electron flux ratios for these models, including the
boost factor, are shown in Figure~\ref{fig19}.

It appears that some of our models do a reasonably good job
of fitting the PAMELA positron data, especially in the case where $\Omega
h^2|_{\mathrm{LSP}}$ lies fairly close to the WMAP value.  
For most models, describing the PAMELA data requires large boost
factors, however this is
also a fairly generic feature of attempts to explain PAMELA and ATIC
data in terms of WIMP annihilation\cite{boost}.  There are however,
many models which give relatively low $\chi^2$ per degree of freedom
in the fit to the data with relatively small boost factors.  We will
study this further in future work \cite{Us DM}.  A study of the
corresponding predictions for the the cosmic ray anti-proton flux is also
underway.

\section{Conclusions}

We have generated a large set of points in parameter space (which we call
``models'') for the 19-parameter CP-conserving pMSSM, where MFV has been
assumed.  We subjected these models to numerous experimental and
theoretical constraints to obtain a set of $\sim 68$~K models which
are consistent with existing data.  We attempted to be
somewhat conservative in our implementation of these constraints; in
particular we only demanded that the relic density of the LSP not
be greater than the measured value of $\Omega h^2$ for non-baryonic
dark matter, rather than assuming that the LSP must account for the
\textit{entire} observed relic density.
 
Examining the properties of the neutralinos in these models, we find
that many are relatively pure gauge eigenstates with Higgsinos being
the most common, followed by Winos.  
The relative prevalence of Higgsino and Wino LSPs leads many
of our models to have a chargino as nLSP, often with a relatively
small mass splitting between this nLSP and the LSP; this has important
consequences in both collider and astroparticle phenomenology.

We find that, in general, the LSP in our models provides a relatively
small ($\sim 4\%$) contribution to the dark matter, however there is a
long tail to this distribution and a substantial number of models for
which the LSP makes up all or most of the dark matter.  Typically
these neutralinos are mostly Binos.  

Examining the signatures of our models in direct and indirect dark
matter detection experiments, we find a wide range of signatures for
both cases.  In particular, we find, not unexpectedly a much larger range of WIMP-nucleon
cross sections than is found in any particular model of SUSY-breaking as can be seen 
by comparing directly with the work in Ref.\cite{Barger:2008qd}.  In fact, as 
these cross sections also enter the regions of parameter space
suggested by non-SUSY models, it appears that the discovery of WIMPs
in direct detection experiments might not be sufficient to determine
the correct model of the underlying physics.  As a first look at the
signatures of these models in indirect detection experiments, we
examined whether our models could explain the PAMELA excess in
the positron to electron ratio at high energies.  We find that there
are models which fit the PAMELA data rather well where the LSP is mostly Bino, 
and some of these 
have significantly smaller boost factors than generally assumed for a
thermal relic.

The study of the pMSSM presents exciting new possibilities for SUSY
phenomenology.  The next few
years will hopefully see important discoveries both in colliders and
in satellite or ground-based astrophysical experiments.  It is
important that we follow the data and not our existing prejudices;
hopefully this sort of relatively model-independent approach to
collider and astrophysical phenomenology can be useful in this regard.

\ack
The authors would like to thank C.~F.~Berger for her
contributions during an earlier stage of this work.
They would also like to thank J.~Conley and D.~Maitre for
computational aid, T.~Tait for discussions related to
this work, and J.~Baglio for his input during the initial stages of
this analysis.  RCC is supported by an NSF Graduate Research
Fellowship.  Work supported in part
by the Department of Energy, Contract DE-AC02-76SF00515.

\clearpage

%
\def\MPL #1 #2 #3 {Mod. Phys. Lett. {\bf#1},\ #2 (#3)}
\def\NPB #1 #2 #3 {Nucl. Phys. {\bf#1},\ #2 (#3)}
\def\PLB #1 #2 #3 {Phys. Lett. {\bf#1},\ #2 (#3)}
\def\PR #1 #2 #3 {Phys. Rep. {\bf#1},\ #2 (#3)}
\def\PRD #1 #2 #3 {Phys. Rev. {\bf#1},\ #2 (#3)}
\def\PRL #1 #2 #3 {Phys. Rev. Lett. {\bf#1},\ #2 (#3)}
\def\RMP #1 #2 #3 {Rev. Mod. Phys. {\bf#1},\ #2 (#3)}
\def\NIM #1 #2 #3 {Nuc. Inst. Meth. {\bf#1},\ #2 (#3)}
\def\ZPC #1 #2 #3 {Z. Phys. {\bf#1},\ #2 (#3)}
\def\EJPC #1 #2 #3 {E. Phys. J. {\bf#1},\ #2 (#3)}
\def\IJMP #1 #2 #3 {Int. J. Mod. Phys. {\bf#1},\ #2 (#3)}
\def\JHEP #1 #2 #3 {J. High En. Phys. {\bf#1},\ #2 (#3)}

\end{document}